\documentclass[12pt,twoside]{article}
\usepackage[mathscr]{eucal}
\usepackage{amsmath,amsfonts,amssymb,amsthm,mathabx,empheq}
\usepackage{times}
\usepackage{pdfsync}
\usepackage{cite}
\usepackage{url}
\usepackage{hyperref}
\usepackage{tensor}
\usepackage{color}

\advance\textheight\topskip
\def\bL{\bar{L}}
\def\bP{\bar{P}}

\def\p{\partial}

\def\bp{\bar \p}
\allowdisplaybreaks[1]
\def\bOmega{\bar\Omega}
\def\bomega{\bar\omega}
\def\bY{\bar Y}
\def\bDelta{\bar \Delta}
\newcommand{\bea}{\begin{eqnarray}}
\newcommand{\eea}{\end{eqnarray}}
\newcommand{\be}{\begin{equation}}
\newcommand{\ee}{\end{equation}}
\newcommand{\bs}{\begin{split}}
\newcommand{\es}{\end{split}}

\newcommand{\dd}{\partial}
\renewcommand{\d}{\partial}

\newcommand{\half}{\frac{1}{2}}

\newcommand{\ffrac}[2]{\raisebox{.5pt}%
  {\footnotesize$\displaystyle\frac{#1}{#2}$}\kern1pt}

\newcommand{\tover}[2]{\ffrac{\partial #1}{\partial #2}}
\newcommand{\dover}[2]{\ffrac{\dd #1}{\dd #2}}

\newcommand{\ddl}[2]{\ffrac{\dd #1}{\dd #2}}

\newcommand{\vddl}[2]{{\ffrac{\delta #1}{\delta #2}}}

\def\cI{\mathcal{I}}
\def\cJ{\mathcal{J}}
\def\cK{\mathcal{K}}
\def\cL{\mathcal{L}}
\def\cM{\mathcal{M}}
\def\cN{\mathcal{N}}
\def\cO{\mathcal{O}}

\def\cY{\mathcal{Y}}

\numberwithin{equation}{section} \makeatletter
\@addtoreset{equation}{section}

\DeclareFontFamily{OT1}{rsfs}{} \DeclareFontShape{OT1}{rsfs}{m}{n}{
<-7> rsfs5 <7-10> rsfs7 <10-> rsfs10}{}
\DeclareMathAlphabet{\mycal}{OT1}{rsfs}{m}{n}

\newcommand*\xbar[1]{%
  \hbox{%
    \vbox{%
      \hrule height 0.5pt 
      \kern0.3ex
      \hbox{%
        \kern-0.0em
        \ensuremath{#1}%
        \kern-0.0em
      }%
    }%
  }%
}

\def\ndelta{\delta\hspace{-0.50em}\slash\hspace{-0.05em} }

\begin{document}

\title{Gravitational current algebra in the context of the
  Newman-Penrose formalism}

\author{Glenn Barnich, Pujian Mao, and Romain Ruzziconi}

\date{}

\def\mytitle{BMS current algebra in the context of the
  Newman-Penrose formalism}

\pagestyle{myheadings} \markboth{\textsc{\small G.~Barnich, P.~Mao, R.~Ruzziconi}}{%
   \textsc{\small Gravitational current algebra in NP formalism}}

\addtolength{\headsep}{4pt}

\begin{centering}

  \vspace{1cm}

  \textbf{\Large{\mytitle}}

  \vspace{1.5cm}

  {\large Glenn Barnich$^{a}$, Pujian Mao$^{b,c}$ and Romain Ruzziconi$^{a}$}

\vspace{.5cm}

\begin{minipage}{.9\textwidth}\small \it  \begin{center}
   ${}^a$ Physique Th\'eorique et Math\'ematique \\ Universit\'e libre de
   Bruxelles and International Solvay Institutes \\ Campus
   Plaine C.P. 231, B-1050 Bruxelles, Belgium
 \end{center}
\end{minipage}

\vspace{.5cm}
\begin{minipage}{.9\textwidth}\small \it  \begin{center}
    ${}^b$ Center for Joint Quantum Studies\\
              and Department of Physics\\
     School of Science, Tianjin University\\
     135 Yaguan Road, Tianjin 300350, P. R. China
 \end{center}
\end{minipage}

\vspace{.5cm}
\begin{minipage}{.9\textwidth}\small \it  \begin{center}
    ${}^c$ Institute of High Energy Physics\\
    and Theoretical Physics Center for Science Facilities \\
    Chinese Academy of Sciences\\ 19B Yuquan Road, Beijing 100049,
    P. R. China
 \end{center}
\end{minipage}

\end{centering}

\vspace{1cm}

\begin{center}
\begin{minipage}{.9\textwidth}
  \textsc{Abstract}. Starting from an action principle adapted to the
  Newman-Penrose formalism, we provide a self-contained derivation of
  BMS current algebra, which includes the generalization of the Bondi
  mass loss formula to all BMS generators. In the spirit of the
  Newman-Penrose approach, infinitesimal diffeomorphisms are expressed
  in terms of four scalars rather than a vector field. In this
  framework, the on-shell closed co-dimension two forms of the
  linearized theory associated with Killing vectors of the background
  are constructed from a standard algorithm. The explicit expression
  for the breaking that occurs when using residual gauge
  transformations instead of exact Killing vectors is worked out and
  related to the presymplectic flux.
 \end{minipage}
\end{center}

\vfill

\thispagestyle{empty}
\newpage
\tableofcontents

\vfill
\newpage

\section{Introduction}
\label{sec:introduction}

The importance of the Bondi mass loss formula
\cite{Bondi:1962px,Sachs1962a} in the context of early research on
gravitational waves has recently been stressed (see
e.g.~\cite{Madler:2016xju,Robinson2017,Kennefick2017}). Since the
(retarded) time translation generator is but one of the generators of
the BMS group \cite{Sachs1962}, a natural problem is to generalize
this formula for all generators (see
e.g.~\cite{Tamburino1966,Geroch1981,Ashtekar1981a,Wald:1999wa}).

Starting from classification results
\cite{Barnich:1994db,Anderson:1996sc,Barnich:2001jy} on conserved
co-dimension $2$ forms in gauge field theories, a BMS charge algebra
\cite{Barnich:2011mi} has been constructed in the metric formulation
in terms of which the non-conservation of BMS charges can be
understood as a particular case.

A local formulation in terms of currents that satisfy a broken
continuity equation then allows one to accommodate both the global
and the local version of the algebra, including
superrotations
\cite{Barnich:2009se,Barnich:2010eb,Barnich:2011ct}. Even though
expressions have been worked out in the metric formulation, they have
been translated to the Newman-Penrose (NP) formalism
\cite{Newman:1961qr,Newman:1962cia,Exton1969} (see
e.g.~\cite{newman:1980xx,Penrose:1984,Chandrasekhar:1985kt,stewart:1991,%
  newman_spin-coefficient_2009} for reviews) in
\cite{Barnich:2011ty,Barnich:2013axa} because the structure of the
results and the interpretation is particularly transparent in this
framework (see also
\cite{He:2014laa,He:2017fsb,Strominger:2017zoo,%
  Godazgar:2018vmm,Alessio:2019cae}
for recent related work.)

The purpose of the present paper is to give a self-contained
derivation, rather than a translation, of BMS current algebra in the
NP formalism by starting from a suitable first order action
principle. The variant we are using here is tensorial (rather than
spinorial as in \cite{Robinson1996,Robinson:1998vf}). It is a first
order action principle of Cartan type that uses as variables
vielbeins, the Lorentz connection in a non-holonomic frame, and a
suitable set of auxiliary fields. In four dimensions, it can be
directly expressed in terms of the quantities of the NP formalism. The
associated Euler-Lagrange equations then impose vanishing of torsion,
the definition of the Riemann tensor in terms of vielbeins and
connection components as well as the Einstein equations, and thus
encode all NP equations.

Related work on conserved quantities in first order formulations of
general relativity includes for instance
\cite{Hehl:1994ue,Julia:1998ys,Julia:2000er,Julia:2002df,%
  Ashtekar:2008jw,JacobsonMohd2015,CorichiRubalcavaVukasinac2014,Lehner:2016vdi,DePaoli:2018erh}.

The paper is organized as follows. We start by reviewing
  the construction of closed co-dimension $2$ forms of the linearized
  theory associated with reducibility parameters of the background,
  including a discussion of the breaking that occurs when using
  residual gauge transformations rather than exact reducibility
  parameters. Standard examples, for instance in the simple case of
  electromagnetism and Yang-Mills theories, have already been
  discussed in \cite{Barnich:2001jy}. Instead, in this paper, we work
  out all details in the case of a generic first order gauge
  theory. In this case, a full understanding of the appropriate
  homotopy operators is not needed because the main statement on
  (non-)conservation of the constructed co-dimension $2$ forms will be
  checked by explicit computation. What makes the paper self-contained
  is that Einstein gravity in the Cartan or the NP formulation are
  gauge theories of first order.

We then turn to the action principle adapted to the NP formalism and
to the construction of the conserved co-dimension $2$ forms in this
context. As an application, we work out the BMS current algebra for
asymptotically flat spacetimes at null infinity. In particular, the
results of \cite{Barnich:2013axa} are generalized to the case of an
arbitrary time-dependent conformal factor.

\section{Closed co-dimension 2 forms in gauge theories}

\subsection{Covariantized Hamiltonian formulations}
\label{sec:covar-hamilt-form}

Let us start by illustrating the general considerations in the case of a first order
theory which depends at most linearly on the derivatives of the
fields,
\begin{equation}
  \label{eq:146}
  L=a^\mu_j\d_\mu\phi^j-h,
\end{equation}
with a generating set of gauge
transformations that depends at most on first order derivatives of the
gauge parameters,
\begin{equation}
  \label{eq:147}
  \delta_f\phi^i= R^i_\alpha[f^\alpha] = R^i_\alpha
  f^\alpha +R^{i\mu}_\alpha\partial_\mu
  f^\alpha,
\end{equation}
and where the derivatives of the fields occur at most linearily in the
term that does not contain derivatives of gauge parameters,
\begin{equation}
  \label{eq:94}
  R^i_\alpha=R^{i0}_\alpha+R^{i\nu}_{j\alpha}\d_\nu\phi^j.
\end{equation}
We thus assume that
$a^\mu_j[x,\phi],
h[x,\phi],R^{i0}_\alpha[x,\phi],R^{i\nu}_{j\alpha}[x,\phi],
R^{i\mu}_\alpha[x,\phi]$ do not depend on derivatives of the
fields. Note that in most applications, and in particular the one of
interest to us here, these functions do not explicitly depend on
$x^\mu$ either and formulas below simplify accordingly.

As the notation indicates, this is a covariantized version of first
order Hamiltonian actions, where $\phi^i$ contains both the canonical
variables and the Lagrange multipliers, while $h$ includes both the
canonical Hamiltonian and the constraints. For instance, for a first
class Hamiltonian system, we have
\begin{equation}
  \label{eq:153}
  L[z,u]=a_A(z)\dot z^A -H(z)-u^a\gamma_a(z).
\end{equation}
Here $z^A$ are the phase-space variables and $a_A(z)$ are
  the components of
  the symplectic potential. In the case of
  Darboux coordinates for instance, $z^A=(q^i,p_j)$ and
  $a_A=(p_1,\dots p_n,0\dots,0)$.  Furthermore, $H$ is the
  Hamiltonian, $\gamma_a$ are the first-class constraints and
  $u^a$ are the associated Lagrange multipliers. The symplectic $2$-form
$\sigma_{AB}=\d_A a_B-\d_Ba_A$ is assumed to be invertible,
$\sigma^{CA}\sigma_{AB}=\delta^C_B$ with associated Poisson bracket
$\{F,G\}=\dover{F}{z^A}\sigma^{AB}\dover{G}{z^B}$ and
\begin{equation}
  \label{eq:154}
  \{\gamma_a,\gamma_b\}=C^c_{ab}(z)\gamma_c,\quad
  \{H,\gamma_a\}=V^b_a(z)\gamma_b.
\end{equation}
For such systems, a generating set of gauge symmetries is given by
\begin{equation}
  \label{eq:155}
  \delta_fz^A=\{z^A,\gamma_a\}f^a,\quad
\delta_fu^a=\dot f^a-C^a_{bc}u^bf^c-V^a_bf^b,
\end{equation}
see e.g.~\cite{Henneaux:1992ig} for more details.

 More generally, by using suitable sets of auxiliary and
  generalized auxiliary fields, the class of gauge theories described
  by \eqref{eq:146} and \eqref{eq:147} is relevant for gravity in the
  standard Cartan formulation or the one adapted to the Newman-Penrose
  formalism discussed below. Indeed, the Cartan Lagrangian is at most
  linear and homogeneous in first order derivatives, and so is the
  Lagrangian adapted to the Newman-Penrose formulation in equation
  \eqref{action NP}. Furthermore, the gauge transformations, both in
  the forms \eqref{eq:15} and \eqref{eq:23} are of the required
  type. Other simpler examples include Chern-Simons theory, which is
  directly of this type, while Yang-Mills theories are of this type
  when using the curvatures as auxiliary fields (see
  e.g.~\cite{Arnowitt:1962aa} for the case of Maxwell's
  theory). Finally, gravity in the Palatini formulation is not of this
type because the transformation of the connection involves second
order derivatives of the vector field characterizing infnitesimal
diffeomorphisms.

For a Lagrangian of the form \eqref{eq:146}, the Euler-Lagrange
operator
$\vddl{L}{\phi^i}=\dover{L}{\phi^i}-\d_\mu\dover{L}{\d_\mu\phi^i}$ is
explicitly given by
\begin{equation}
  \label{eq:148}
  \vddl{L}{\phi^i}=\sigma^{\mu}_{ij}\d_\mu\phi^j-\d_ih
  -\dover{}{x^\mu} a^\mu_i,\quad \sigma^\mu_{ij}=\d_i a^\mu_j-\d_j
  a^\mu_i \Longrightarrow \d_{[i} \sigma^\mu_{jk]}=0,
\end{equation}
where $\d_i=\dover{}{\phi^i}$,  while
  $\d_\mu=\dover{}{x^\mu}+\d_\mu\phi^i\dover{}{\phi^i}+\dots$. For a
global symmetry $\delta_Q \phi^i = Q^i$, it follows that
$\delta_Q L=\partial_\mu b^\mu_Q$ and the usual integrations by parts
argument to go to the form
\begin{equation}
  \label{eq:10}
  Q^i\vddl{L}{\phi^i}=\partial_\mu j^\mu_Q,
\end{equation}
yields  a canonical representative for the associated Noether
current given by
\begin{equation}
  \label{eq:6}
  j^\mu_Q=b^\mu_Q-\dover{L}{\partial_\mu \phi^i}Q^i.
\end{equation}

When inserting the generating set of gauge symmetries
  given in \eqref{eq:147} into
\eqref{eq:10}, one obtains on the one hand
\begin{equation}
R^i_\alpha[f^\alpha]
\vddl{L}{\phi^i}=\partial_\mu j^\mu_f.\label{eq:104}
\end{equation}
Doing integrations by parts on the expression on the left hand side so
as to make the undifferentiated gauge parameters appear, one obtains
on the other hand
\begin{equation}
  \label{eq:60}
R^i_\alpha[f^\alpha]
\vddl{L}{\phi^i}=  f^\alpha [R^i_\alpha \vddl{L}{\phi^i}-\partial_\mu (R^{i\mu}_\alpha
  \vddl{L}{\phi^i})]+ \partial_\mu S^\mu_f,\quad S^\mu_f=f^\alpha R^{i\mu}_\alpha
  \vddl{L}{\phi^i}.
\end{equation}
Subtracting these two equations gives
\begin{equation}
  \label{eq:106}
f^\alpha [R^i_\alpha \vddl{L}{\phi^i}-\partial_\mu (R^{i\mu}_\alpha
  \vddl{L}{\phi^i})]=\d_\mu(j^\mu_f-S^\mu_f).
\end{equation}
Since this is an off-shell identity that has to hold for all
$f^\alpha[x]$,
one concludes not only that the Noether
identities
\begin{equation}
  \label{eq:149}
  R^i_\alpha\vddl{L}{\phi^i}-\d_\mu(R^{i\mu}_\alpha\vddl{L}{\phi^i})=0,
\end{equation}
hold, but also that
\begin{equation}
\d_\mu(j^\mu_f-S^\mu_f)=0.\label{eq:103}
\end{equation}
There are two ways of seeing this. Either one integrates equation
\eqref{eq:106} with gauge parameters that vanish on the boundary. One
then uses Stokes' theorem and the fact that the parameters are still
arbitrary in the bulk to conclude that the Noether identities must
hold, and then that the right hand side must be zero as
well. Alternatively, in a more algebraic approach, the fact that the
gauge parameters are arbitrary allows one to consider them as new
fields with respect to which one can take Euler-Lagrange
derivatives. If Euler-Lagrange derivatives with respect to the gauge
parameters are applied to equation \eqref{eq:106}, the right hand side
vanishes identically because Euler-Lagrange derivatives annihilate
total divergences. One then remains with the Noether identities as
the result of the application of Euler-Lagrange derivatives with
respect to the gauge parameters on the left hand side.

When using these Noether identities, it then also follows from
\eqref{eq:60} that
\begin{equation}
  \label{eq:107}
  R^i_\alpha[f^\alpha]\vddl{L}{\phi^i}=\d_\mu S^\mu_f.
\end{equation}
This means that $S^\mu_f$ is a representative for the Noether current
associated to gauge symmetries that is trivial in the sense that it
vanishes on-shell. This is Noether's second theorem.  Furthermore, it
also follows from the so-called algebraic Poincar\'e lemma (or in
other words, the local exactness of the horizontal part of the
variational bicomplex in form degrees less than the spacetime
dimension, see e.g.~\cite{Olver:1993,Andersonbook}) applied to
\eqref{eq:103} that every other representative $j^\mu_f$ differs from
$S^\mu_f$ at most by the divergence of an arbitrary superpotential
$\d_\nu \eta^{[\mu\nu]}_f$ (in the absence of non-trivial topology).

Equation \eqref{eq:107} provides a way to associate
(lower-dimensional) conservation laws with particular gauge
symmetries: indeed, for reducibility parameters, i.e., gauge
parameters $\bar f^\alpha$ that satisfy
\begin{equation}
  \label{eq:108}
  R^i_\alpha[\bar f^\alpha]=0,
\end{equation}
the local exactness of the horizontal part of the variational
bicomplex in form degree $n-1$ then implies the existence of a
superpotential $k^{[\mu\nu]}_{\bar f}$ that is constructed out of the
fields and a finite number of their derivatives such that
\begin{equation}
  \label{eq:109}
  \d_\nu k^{[\mu\nu]}_{\bar f}=S^\mu_{\bar f}.
\end{equation}
Since the right hand side vanishes on all solutions of the field
equations, this is a conservation law whose associated co-dimension
$2$-form should be integrated over co-dimension 2 surfaces.

The point is that equation \eqref{eq:108} does not in general admit
non-trivial solutions in truly interacting gauge theories. Indeed, in
the case of semi-simple Yang-Mills theories or general relativity in
metric formulation, this equation reads explicitly
\begin{equation}
  \label{eq:110}
  D_\mu\bar \epsilon^a=0,\quad \cL_{\bar \xi} g_{\mu\nu}=0.
\end{equation}
Since this equation has to hold for all gauge potentials or metrics,
it turns out however that there are no non trivial solutions to these
equations. In the latter case, this is because a generic metric does
not have Killing vectors.

This is where the linearized theory comes in. Consider the
linearization around a particular background solution $\bar \phi^i$,
with $\phi^i=\bar\phi^i+\varphi^i$ and where $\varphi^i$ denotes the
fluctuations.

Two facts about linearized gauge theories around a solution are
important. The first is that the linearized equations of motion are
variational and derive from the quadratic part
$L^{(2)}[\bar\phi,\varphi]$ of the Lagrangian in the fluctuations. The
second is that gauge symmetries of this linearized gauge theory are
obtained by evaluation of the gauge transformations of the full theory
in the background,
$\delta_f\varphi^i= R^i_\alpha[f^\alpha]|_{\phi=\bar\phi}$. This
follows by expanding equation \eqref{eq:107} to first order in the
fluctuations. Furthermore, by expanding this equation to second order
and using reducbility parameters, it follows that the transformations
$\delta_{\bar f}\varphi^i=R^{(1)i}_\alpha(\bar f^\alpha)$ define global
symmetries of the linearized theory.

In the case of general relativity in the metric formulation, this is
the statement that the gauge symmetries of the quadratic Lagrangian
for fluctuations around a fixed background are given by
$\delta_\xi h_{\mu\nu}=\cL_\xi \bar g_{\mu\nu}$. This then means that
there are as many conserved co-dimension 2 forms as there are Killing
vectors of the background solution. In the simplest case of
linearization around flat space, one recovers in this way the gauge
symmetries $\delta_\xi h_{\mu\nu} =\d_\mu \xi_\nu+\d_\nu\xi_\mu$ of
the Pauli-Fierz Lagrangian. The reducibility parameters are the
Killing vectors of the background, and thus given
$\bar \xi_\mu=a_\mu+\omega_{[\mu\nu]}x^\nu$ in the case of Pauli-Fierz
theory for instance. In this case, the associated global symmetries
$\delta_{\bar \xi}h_{\mu\nu}=\cL_{\bar\xi}h_{\mu\nu}$ describe the
invariance of Pauli-Fierz theory under Poincar\'e transformations.

The remaing problem is then to explicitly construct
$k^{[\mu\nu]}_{\bar f}$ of the linearized theory out of the weakly
vanishing Noether current $S^\mu_{\bar f}$ of the full theory. This is
done with a suitable homotopy operator of the variational
bicomplex. The point is that this operator is quite complicated in
case $S^\mu_{\bar f}$ contains derivatives of the fields of higher
order than one because it involves higher order Euler-Lagrange
derivatives.

In a first order gauge theory however, $S^\mu_{\bar f}$ depends at
most linearily on first order derivatives by assumption. In this case,
the homotopy operator becomes quite simple and the expression of the
superpotential simplifies to
\begin{equation}
  \label{eq:111}
  k^{[\mu\nu]}_{\bar f}=\half
  \varphi^j\dover{}{\d_{\nu}\phi^j}S^\mu_{\bar f}-(\mu\leftrightarrow
  \nu).
\end{equation}
When using the explicit expression for $S^\mu_{f}$ in \eqref{eq:60}
and for the left hand sides of the field equations in \eqref{eq:148},
this gives
\begin{equation}
    \label{eq:81}
k^{[\mu\nu]}_f=
  R^{i[\mu}_\alpha\sigma^{\nu]}_{ij}\varphi^jf^\alpha.
\end{equation}
In this case, one can avoid a detailed discussion of the homotopy
operator and its properties and limit oneself to simply verify that
this superpotential is indeed conserved on all solutions of the
linearized field equations around a background solution of the full
theory, when using reducibility parameters. Furthermore, one may
explicitly work
out the breaking of this conservation law in case one does not use
such reducibility parameters. For this purpose, one introduces the
presymplectic current,
\begin{equation}
  \label{eq:83}
  W^\mu_{{\delta\cL}/{\delta\phi}}[\varphi_1,\varphi_2]=-\sigma_{ij}^\mu\varphi^i_1\varphi^j_2.
\end{equation}
By using the detailed form of the Noether identities, one may then
check by a direct computation that
\begin{equation}
  \label{eq:85}
  \d_\nu
  k^{[\mu\nu]}_f=-W^\mu_{{\delta\cL}/{\delta\phi}}[\varphi,R_\alpha[f^\alpha]] \quad {\rm when}\quad
    \vddl{L}{\phi^i}=0=\vddl{L^{(2)}[\varphi,\phi]}{\varphi^i}.
  \end{equation}
  All details are provided in Appendix \ref{sec:non-cons-codim}.
    This means that these superpotentials are indeed conserved on all
    solutions of the linearized equations of motion around a given
    background solution $\phi$ when using reducibility parameters
    $\bar f^\alpha$ satisfying \eqref{eq:108}. It also means that
    non-conservation in case one uses more general gauge parameters is
    controlled by the symplectic flux
    $W^\mu_{{\delta\cL}/{\delta\phi}}[\varphi,R_\alpha[f^\alpha]]$.

In terms of 1-forms $dx^\mu, d_V\phi^i,\d_\mu d_V\phi^i,\dots$
generating the variational bicomplex (see
e.g. \cite{Andersonbook,Olver:1993}), one may write
\begin{equation}
  \label{eq:157}
\boxed{ \star k_f=
  R^{i\mu}_\alpha\sigma^{\nu}_{ij}\varphi^jf^\alpha \half \star (
  dx_\mu dx_\nu)},
\end{equation}
where
\begin{equation}
    \label{eq:159}
    d(\star k_f)=-W_{{\delta\cL}/{\delta\phi}}[\varphi,R_f]\quad {\rm when}\quad
    \vddl{L}{\phi^i}=0=\vddl{L^{(2)}[\varphi,\phi]}{\varphi^i},
  \end{equation}
  with
  \begin{equation}
    W_{{\delta\cL}/{\delta\phi}}=-\half
  \sigma^\mu_{ij}d_V\phi^id_V\phi^j
  \star(dx_\mu),
  \end{equation}
  $d=  dx^\mu \partial_\mu = dx^\mu(\dover{}{x^\mu}+\partial_\mu\phi^i\dover{}{\phi^i}+\partial_\mu
  d_V\phi^i\dover{}{d_V\phi^i}\dots)$,  where the
    vertical exterior derivative $d_V
    \phi^i$ can be considered as the dual $1$-form to an infinitesimal
    field variation that commutes with the total derivative, and
  we
  have used the conventions for the Hodge dual of Appendix
  \ref{sec:conventions-forms}, with $*e^a=\delta^a_\mu dx^\mu$.

Defining the presymplectic 1 form potential through
\begin{equation}
  \label{eq:158}
 a=a^\mu_id_V\phi^i \star
  (dx_\mu),
\end{equation}
we have
\begin{equation}
  \label{eq:92}
 \Omega_{\cL}=d_Va=-W_{{\delta\cL}/{\delta\phi}}
\end{equation}
and equation \eqref{eq:157} can be expressed in terms of $a$ as
\begin{equation}
  \label{eq:161}
 \star k_f=\half(\varphi^j\frac{\d}{\d d_V\phi^j}) (f^\alpha
 R^{i\mu}_\alpha\frac{\d}{\d d_V\phi^i})\frac{\d}{\d dx^\mu}d_V a.
\end{equation}

For instance, in the standard Cartan formulation, the variables are
the vielbein and the Lorentz connnection,
$\phi^i=({e_a}^\mu,{\Gamma^{ab}}_\mu)$, with
\begin{equation}
  \begin{split}
\delta_{\xi,\omega} {e_a}^\lambda=\xi^\rho\d_\rho
{e_a}^\lambda-\d_\rho \xi^\lambda
{e_a}^\rho+{\omega_a}^b{e_b}^\lambda,\\
\delta_{\xi,\omega}
{\Gamma^{ab}}_\mu=-D_\mu\omega^{ab}+\xi^\rho\d_\rho{\Gamma^{ab}}_\mu+\d_\mu\xi^\rho{\Gamma^{ab}}_\rho. \label{eq:162}
\end{split}
\end{equation}
From
\begin{equation}
L_C=\mathbf e\,({R^{ab}}_{\mu\nu}{e_a}^\mu
{e_b}^\nu-2\Lambda^C),\label{eq:163}
\end{equation}
it then follows that
\begin{equation}
  \label{eq:160}
a=  2\mathbf e\,{e_a}^\mu{e_b}^\rho
  d_V{\Gamma^{ab}}_\rho \star (dx_\mu).
\end{equation}

Applying equation \eqref{eq:161} for the construction of $\star k_f$
then yields quickly equation (3.49) of \cite{Barnich:2016rwk} when
using that $(d^{n-2}x)_{\mu\nu}=\half\star (dx_\mu dx_\nu)$.

We now turn to a more systematic discussion of the construction of
these superpotentials, with no assumptions on the number of
derivatives except for locality, i.e., the requirement that all
functions contain a finite number of derivatives. This discussion may
be skipped if one is merely interested in the application to the NP
formalism.

\subsection{General construction}
\label{sec:gener-cons}

We continue to use the collective notation $\phi^i$ for all fields and
$f^\alpha[x,\phi]$ for the gauge parameters of the theory. The latter
may depend on the fields and (a finite number of) their
derivatives. Infinitesimal gauge transformations are written as
$\delta_f\phi^i=  R^i_\alpha[f^\alpha] \equiv R^i_f[\phi]$, where the second
notation is used when we want to emphasize the dependence on the
fields. More explicitly, they involve field dependent
operators acting on the gauge parameters,
\begin{equation}
R^i_\alpha[f^\alpha]\equiv R^i_f[\phi] =R^i_\alpha[x,\phi]
f^\alpha+R^{i\mu}_\alpha[x,\phi]\partial_\mu
f^\alpha+\dots\label{eq:21},
\end{equation}
which may now depend on the fields and a finite number of their
derivatives. Furthermore, there may be terms with higher order
derivatives on the gauge parameters.

In the case of general relativity (or higher derivative gravitational
theories) in metric formulation for instance, the fields $\phi^i$
correspond to the metric components $g_{\mu\nu}$, while the gauge
parameters are vector fields $\xi^\mu[x,g]$, with
$\delta_\xi g_{\mu\nu}=-\mathcal L_\xi g_{\mu\nu}$.

Isolating the undifferentiated gauge parameters in the contraction of
the gauge transformations with the Euler-Lagrange derivatives of the
Lagrangian and keeping the total derivative terms gives rise to
\begin{equation}
  \label{eq:43}
  \delta_{f}\phi^i\vddl{\mathcal L}{\phi^i}= f^\alpha
  R^{+i}_\alpha(\vddl{\mathcal L}{\phi^i})+
  d \star S_{f},
\end{equation}
where ${\mathcal L}=\star L$ is the Lagrangian $n$-form and
\begin{equation}
  \label{eq:7}
  R^{+i}_\alpha(\vddl{\mathcal L}{\phi^i})=R^i_\alpha \vddl{\mathcal
    L}{\phi^i} -\d_\mu(R^{i\mu}_\alpha\vddl{\mathcal
    L}{\phi^i})+\dots,\quad \star S_f=f^\alpha R^{i\mu}_\alpha
  \tover{}{dx^\mu}\vddl{\cL}{\phi^i}+\dots.
\end{equation}
Because the $f^\alpha$ are arbitrary, one
deduces on the one hand the Noether identities
\begin{equation}
  \label{eq:71}
  R^{+i}_\alpha(\vddl{\mathcal L}{\phi^i})= 0,
\end{equation}
and on the other hand that
the $1$ form $S_{f}$ encodes the on-shell vanishing Noether
current associated to gauge symmetries,
\begin{equation}
  \label{eq:57}
  S_f\approx 0.
\end{equation}

Closed co-dimension $2$ forms are then constructed as follows. Consider
an infinitesimal field variation
\begin{equation}
\delta^V=\varphi^i\ddl{}{\phi^i}
+\d_\mu\varphi^i\ddl{}{\d_\mu\phi^i}+\dots\label{eq:42}.
\end{equation}
Acting on \eqref{eq:43} (with \eqref{eq:71} taken into account) gives
\begin{equation}
  \label{eq:44}
  \delta^V(\delta_{f}\phi^i)\vddl{\mathcal
    L}{\phi^i}+\delta_{f}\phi^i
  \delta^V(\vddl{\mathcal
    L}{\phi^i})=d[\delta^V(\star S_{f})].
\end{equation}
Let $\mathcal L^{(2)}[\phi,\varphi]$ denote the quadratic terms in an
expansion of $\mathcal L[\phi+\varphi]$ in $\varphi^i$ and their
derivatives. The equations of motion of the linearized theory are
the Euler-Lagrange equations for $\mathcal L^{(2)}$ since
\begin{equation}
  \label{eq:47}
  \delta^V\vddl{\mathcal L}{\phi^i}=\vddl{\mathcal L^{(2)}}{\varphi^i}.
\end{equation}
If the expansion is done around any solution $\bar\phi^i$ of
the full theory associated to $\mathcal L[\phi]$,
\begin{equation}
  \label{eq:72}
  \vddl{\mathcal L}{\phi^i}|_{\bar\phi}=0,
\end{equation}
it follows in particular that, when evaluated on solutions of the full
theory, $\delta^V S_{f}$
vanishes on all solutions to the linearized equations of motion,
\begin{equation}
  \label{eq:66}
  (\delta^V S_f)|_{\bar\phi}\approx_{\rm lin}0.
\end{equation}
If furthermore the parameters $\bar f^\alpha$ are ``reducibility
parameters'' of this solution,
\begin{equation}
  \label{eq:50}
  \delta_{\bar f}\phi^i\approx_{\rm full} 0,
\end{equation}
the left hand side of \eqref{eq:44} vanishes. When using that closed
co-dimension $1$ forms are exact in this context (see Appendix
\ref{sec:homot-oper-euler}), one finds
\begin{equation}
  \label{eq:51}
  \delta^V(\star S_{f})|_{\bar\phi,\bar f}=d(\star k'_{f})|_{\bar\phi,\bar f},\quad
  \star k'_{f}[\varphi,\phi]=I^{n-1}_\varphi(\star
  S_{f}),
\end{equation}
with $I^{n-1}_\varphi$ given in \eqref{eq:64} and \eqref{eq:63}.
This equation means that $\star k'_{\bar f}[\bar\varphi,\bar \phi]$ are
local co-dimension $2$ forms that are closed if (i) $\bar\phi$ is a
solution to the full theory, (ii) $\bar f$ are reducibility
parameters of this solution, and (iii) $\bar\varphi$ are solutions
of the linearized theory around $\bar\phi$,
\begin{equation}
  \label{eq:52}
d\big(\star k'_{\bar f}[\bar\varphi,\bar\phi])=0.
\end{equation}
Finally, note that these co-dimension $2$ forms may be computed as if
the gauge parameters were field independent,
\begin{equation}
  \label{eq:73}
   \star k_f=I^{n-1}_\varphi[\star S_{f(x)}]|_{f(x)=f},
 \end{equation}
 because, as shown in Appendix \ref{sec:homot-oper-euler},
 \begin{equation}
\star k_{\bar f}[\bar\phi,\varphi]=\star k'_{\bar
  f}[\bar\phi,\varphi]\label{eq:74}
\end{equation}
for all solutions $\bar\phi^i$.

Applying \eqref{eq:73} together with the homotopy formula in the form of \eqref{eq:63}
with $*e^a=\delta^a_\mu dx^\mu$ to the weakly vanishing current
\eqref{eq:60} of a
covariantized first order Hamiltionian theory then yields the simple
expression \eqref{eq:157}.

\subsection{Application in different contexts}
\label{sec:appl-diff-situ}

\subsubsection{Linearized gauge theories}
\label{sec:line-gauge-theor}

Different cases can be considered. The first is to fix a background
solution $\bar\phi$ with its reducibility parameters $\bar f$, for
instance maximally symmetric backgrounds in general relativity with
its Killing vectors. The second is to fix a priori a set of
reducibility parameters and to restrict to classes of solutions
$\bar\phi$ that admit these reducibility parameters (stationnary
and/or axisymmetric backgrounds in general relativity). In both cases
$k_{\bar f}[\varphi,\bar\phi]$ are co-dimension $2$ forms that are
closed for all solutions $\varphi$ of the linearized theory,
\begin{equation}
  \label{eq:58}
  \boxed{d(\star k_{\bar f}[\varphi,\bar\phi])\approx_{\rm lin}0}.
\end{equation}

Equivalent closed co-dimension $2$ forms have been derived by a
variety of methods (in the case of diffeomorphism invariant theories,
see e.g.~\cite{Wald:1990ic,Lee:1990nz,Iyer:1994ys,Iyer:1995kg}) and
used to provide a derivation of the first law of black hole mechanics
\cite{Bardeen:1973gs} valid for arbitrary perturbations.

An advantage of the approach of \cite{Barnich:1994db} (see also
\cite{Anderson:1996sc} and
\cite{Barnich:2000zw,Barnich:2001jy,Barnich:2003xg,Barnich:2007bf,Barnich:2016rwk}
for further developments) is that it can be used for any gauge theory
and that there is complete control on the number of solutions and on
the ambiguities of the construction: under suitable regularity
conditions, the $\star k_{\bar f}[\varphi,\bar\phi]$ associated to
distinct equivalence classes of possibly field dependent
$\bar f^\alpha$'s satisfying \eqref{eq:50}, with two sets of
reducibility parameters considered equivalent if their difference are
reducibility parameters that vanish on-shell,
$\bar f^\alpha\approx 0$, exhaust the local co-dimension $2$ forms
that are closed on all solutions to the linearized equations, up to
trivial ones. The latter correspond to local co-dimension $2$ forms
that are $d$ exact or vanish on all solutions of the linearized
theory. Furthermore, the equivalence classes do not depend on the
formulation, in the sense that they are invariant under elimination or
introduction of (generalized) auxiliary fields, which allows one to
directly connect results in the Cartan and metric formulations for
instance.

In the linearized theory, reducibility parameters give rise in
addition to global symmetries with standard Noether currents, i.e.,
closed forms of co-dimension $1$. The gauge algebra of the full theory
then induces Lie algebra structures on reducibility parameters and on
the closed co-dimension $1$ and $2$ forms (see e.g.~section 7.4 of
\cite{Barnich:2001jy} or Proposition 4 and Corollary 5 of
\cite{Barnich:2007bf}).

The method can usually not be directly used to construct closed
co-dimension $2$ forms for generic background solutions in interacting
theories, such as semi-simple Yang-Mills theories or general
relativity in spacetime dimension greater than $3$ because the
equation determining the reducibility parameters $\bar f^\alpha$ in
\eqref{eq:50} admits no non-trivial solutions, not even when
allowing the gauge parameters to be field dependent. This is where
asymptotic considerations come in. We will not discuss this here, but
take a slightly different viewpoint below.

\subsubsection{Residual gauge transformations and breaking}
\label{sec:residual-symmetries}

Another important case, that is the one that is relevant for us here,
is to fix from the outset a sub-class of solutions $\hat\phi$. This
can be done not only through gauge fixing conditions, like asking for
certain components of the metric or the vielbein/spin-connection to
vanish, but also through fall-off conditions. The role of these
conditions is then to restrict the arbitrary functions that appear in
the general solution to the equations of motion. These conditions may
be imposed anywhere, and are not limited to conditions imposed at
``infinity''. We assume that these conditions are such that one may
find the general solution to the equations of motion in terms of
functions $a^A(x)$ (which could reduce to constants),
$\hat \phi=\hat \phi[x,a]$. The functions $a^A(x)$ thus parametrize
this solution space, and infinitesimal variations of these functions
lead to tangent vectors of this solution space, i.e., to perturbations
$\hat\phi$ that are solutions of the linearized equations of motion.

One is then interested in (infinitesimal) gauge transformations that
preserve this class of solutions. They are determined by asking that
the gauge transformations preserve the conditions fixing the solution
space.  We assume that this constrains the parameters to depend on
arbitrary functions $b^M(x)$, $\hat f=\hat f[x,b;a]$.  The associated
infinitesimal gauge transformations, loosely referred to as residual
gauge transformations below, no longer satisfy
$\delta_{\hat f}\phi^i|_{\hat\phi}=R^i_{\hat f}[\hat\phi]=0$. Instead
they describe particular tangent vectors to the subspace of solutions
and correspond to particular variations $\delta_{b}a^A$. This action
of residual gauge transformations on solution space is thus determined
through
\begin{equation}
  \label{eq:156}
  R^i_{\hat f}[\hat\phi]=-\delta_b\hat\phi^i[x,a].
\end{equation}
As a
consequence, the co-dimension $2$ forms are no longer closed. There is
however precise control on the breaking, i.e., on non-conservation: it
is proportional to $R_{\hat f}[\hat\phi]$, and furthermore, it follows
from \eqref{eq:44} and \eqref{eq:68} that
\begin{equation}
  \label{eq:65}
  \boxed{d(\star k_{\hat f}[\hat\varphi,\hat\phi])=\star
    b[\hat\varphi,R_{\hat f},\hat\phi]},\quad \star b[\varphi,R_{f};\phi]=-
  \cI^n_\varphi(R_f^i\delta^V\vddl{\cL}{\phi^i}),
\end{equation}
with $\cI^n_\varphi$ defined in \eqref{eq:54}. This allows one for
instance to control both non-conservation in (retarded-)time or the
radial dependence of charges by using
\begin{equation}
  \label{eq:144}
  \int_{\partial \cN^{n-1}} \star k_{\hat
    f}[\hat\varphi,\hat\phi]=\int_{\cN^{n-1}} \star
  b[\hat\varphi,R_{\hat f};\hat\phi].
\end{equation}
More concretely for instance, in terms of coordinates $u,r,y^A$, with
$y^A$ parametrizing an $n-2$ dimensional sphere, the time-dependence of
the charges
\begin{equation}
(\ndelta Q_{\hat f})[\hat\varphi,\hat\phi]=\int d^{n-2}\Omega\ k^{0r}_{\hat
    f}[\hat\varphi,\hat\phi]\label{eq:89}
\end{equation}
is controlled by
\begin{equation}
\d_0 k^{r0}_{\hat
    f}[\hat\varphi,\hat\phi]+\d_A k^{rA}_{\hat
    f}[\hat\varphi,\hat\phi]=b^r[\hat\varphi,R_{\hat
    f};\hat\phi]\label{eq:87}
\end{equation}
while the radial dependence is controlled by
\begin{equation}
  \label{eq:88}
\d_r k^{0r}_{\hat
    f}[\hat\varphi,\hat\phi]+\d_A k^{0A}_{\hat
    f}[\hat\varphi,\hat\phi]=b^0[\hat\varphi,R_{\hat
    f};\hat\phi],
\end{equation}
where the second terms on the left hand side vanish when integrating
over the sphere.

Let us recall some properties of $\star b[R_{f},\varphi;\phi]$. A non
trivial property, which follows because the contracting homotopy is
applied to an Euler-Lagrange derivative, is skew-symmetry in the
exchange of the infinitesimal gauge transformation and the field
variation,
\begin{equation}
  \label{eq:139}
  \star b[\varphi,R_{f};\phi]=-\star b[R_{f},\varphi;\phi].
\end{equation}
For first order theories, this has been shown in \cite{Julia:2002df},
while for general theories, the proof follows from that of
proposition 13, and more precisely from equation (A59), of
\cite{Barnich:2007bf}.

The link to covariant phase space methods is as follows. If we
consider anti-commuting infinitesimal field variations, $\delta^V\phi^i$,
and the associated vertical differential
$d_V=d_V\phi^i\tover{}{\phi^i}+\partial_\mu
d_V\phi^i\tover{}{\d_\mu\phi^i}+\dots$ (see
e.g. \cite{Andersonbook,Olver:1993} for more details), one may define
two $(n-1,2)$ forms. The first is the standard presysmplectic one of
variational calculus,
\begin{equation}
  \label{eq:140}
  \Omega_{\cL}=d_V I^n_{d_V\phi}\cL,\quad d_V\Omega_{\cL}=0, \quad d
  \Omega_{\cL}=d_V\phi^id_V\vddl{\cL}{\phi^i},
\end{equation}
with all wedge products omitted. When using the main property of the
homotopy operators (cf. equation (A30) of \cite{Barnich:2007bf}), it
follows that for $\cL'=\cL+d\theta^{n-1}$,
\begin{equation}
  \label{eq:145}
  \Omega_{\cL'}=\Omega_{\cL}+d(d_V I^{n-1}_{d_V\phi}\theta^{n-1}).
\end{equation}
Note that the ambiguity does not affect the presymplectic form in the
restricted class of first order Lagrangians, where $\theta^{n-1}$
may depend on undifferentiated fields only, so that the last term in
the previous equation vanishes.

The second ``invariant'' presymplectic $(n-1,2)$ form
\cite{Julia:2002df,Barnich:2007bf} is defined through,
\begin{equation}
  \label{eq:141}
  W_{{\delta\cL}/{\delta\phi}}=-\half
  I^n_{d_V\phi}\big(d_V\phi^i\vddl{\cL}{\phi^i}\big).
\end{equation}
It depends only on the Euler-Lagrange derivatives of the Lagrangian and
is thus free of the ambiguity related to adding a total derivative to the
Lagrangian.

By using proposition 13 of appendix A5 of \cite{Barnich:2007bf}, it
follows that the breaking is determined by the invariant
presymplectic form according to
\begin{equation}
  \label{eq:123}
  \boxed{\star b[R_f,\varphi;\phi]=W_{{\delta\cL}/{\delta\phi}}[\varphi,R_f]}.
\end{equation}
Up to a sign, both $(n-1,2)$ forms differ by the exterior derivative
of an $(n-2,2)$-form,
\begin{equation}
  \label{eq:142}
  -W_{{\delta\cL}/{\delta\phi}}=\Omega_{\cL}+dE_{\cL},\quad
  E_{\cL}=\half I^{n-1}_{d_V\phi}I^n_{d_V\phi}\cL.
\end{equation}
In the particular case of first order theories where $\cL$ depends at most
linearly on the first order derivatives, the explicit expression in
terms of homotopy operators shows that
\begin{equation}
  \label{eq:143}
  E_{\cL}=0.
\end{equation}

It may happen that
$W_{{\delta\cL}/{\delta\phi}}[\hat\varphi,R_{\hat f}]|_{\hat\phi}$
vanishes. Examples include for instance asymptotically flat or anti-de
Sitter spacetimes in three dimensions in Fefferman-Graham or BMS gauge
with fixed conformal factor. In this case, the co-dimension $2$ forms
are closed for all residual gauge transformations, they are conserved
and $r$-independent (see for instance
\cite{Compere:2014cna,Compere:2015knw}), so that they may be computed
at any finite $r$. It also follows from \eqref{eq:88} and
\eqref{eq:123} that subleading charges recently considered for
instance in \cite{Conde:2016csj,Conde:2016rom,Godazgar:2018vmm} are controlled by
$W^0_{{\delta\cL}/{\delta\phi}}[\hat\varphi,R_{\hat f}]|_{\hat\phi}$.

\subsection{Integrability and algebra}
\label{sec:integr-algebra}

When $E_{\cL}$ vanishes, it has been shown in \cite{Barnich:2007bf}
that integrability of charges
\begin{equation}
(\ndelta Q_{\hat
  f})[\hat\varphi,\hat\phi]=\delta Q_{\hat f}[\hat\phi],\label{eq:90}
\end{equation}
implies that there is well-defined algebra of charges obtained by
acting with residual gauge transformations,
$-\delta_{\hat f_2} Q_{\hat f_1}$. In the non-integrable case, this
action has to be modified by suitably taking the non-integrable part
of the charges into account. Even though it would be desirably to have
a derivation from first principles of this modified charge or current
algebra, this is not the objective of this paper. At this stage, we
just refer to the discussion in section 3.2 of \cite{Barnich:2011mi}
or in sections 4.2 and 4.3 of \cite{Barnich:2013axa} and concentrate on
a self-contained derivation of $\star k_f$ and of the breaking in the
context of the NP formalism.

\section{Cartan formalism in non-holonomic frame}
\label{sec:newm-penr-form}

\subsection{Connection, torsion and curvature}
\label{sec:cartannonholo}

Besides the non-holonomic frame and the peudo-Riemmanian metric
discussed in Appendix \ref{sec:conventions-forms}, we now furthermore
assume that there exists an affine connection whose components in the
non-holonomic basis are
\begin{equation}
D_a e_b={\Gamma^c}_{ba}e_c\iff {\Gamma^a}_{bc}= {e^a}_\mu D_c
{e_b}^\mu =-{e_b}^\mu D_c {e^a}_\mu\label{eq:55}.
\end{equation}
It follows that torsion and curvature components are given by
(see Appendix \ref{sec:conventions-forms} for notations
  and conventions)
\begin{equation}
  \begin{split}
  \label{eq:12}
&  {T^a}_{bc}=2{\Gamma^a}_{[cb]}+{D^a}_{cb}=2({\Gamma^a}_{[cb]}+{d^a}_{[cb]}),\\
& {R^a}_{bcd}=\d_c
  {\Gamma^a}_{bd}-\d_d {\Gamma^a}_{bc}
  +{\Gamma^a}_{fc}{\Gamma^f}_{bd}-{\Gamma^a}_{fd}{\Gamma^f}_{bc}-{\Gamma^a}_{bf}{D^f}_{cd},
\end{split}
\end{equation}
and
\begin{equation}
  \label{eq:20}
  [D_a,D_b]v_c=-{R^d}_{cab}v_d-{T^d}_{ab}D_dv_c.
\end{equation}
The reason torsion has to be included in the discussion is that it
will vanish only for solutions of the field equations. In off-shell considerations based on
the action principle proposed below, non-vanishing torsion terms have
to be taken into account.

The Bianchi identities are
\begin{equation}
  \label{eq:24}
  {R^a}_{[bcd]}=D_{[b}{T^a}_{cd]}+{T^a}_{f[b}{T^f}_{cd]},\quad
D_{[a}{R^f}_{|b|cd]}=-{R^f}_{bh[a}{T^h}_{cd]},
\end{equation}
where a bar encloses indices that are not involved in the (anti)
symmetrization. The Ricci tensor is defined by
${R}_{ab}={R^{c}}_{acb}$, while $S_{ab}={R^c}_{cab}$. Various
contraction of the Bianchi identities give
\begin{equation}
  \label{eq:27}
  {R}_{ab}-{R}_{ba}=S_{ab}-D_c {T^c}_{ab}
-2D_{[a} {T^c}_{b]c}-{T^c}_{fc}{T^f}_{ab},
\end{equation}
\begin{equation}
2D_{[a}{R}_{|b|c]}+D_d{R^d}_{bca}={R}_{bd}{T^d}_{ca}
-2{R^d}_{b[a|f|}{T^f}_{c]d}, \label{eq:28a}
\end{equation}
\begin{equation}
  \label{eq:28b}
  D_{[a}S_{bc]}=-S_{d[a}{T^d}_{bc]}.
\end{equation}

The connection is assumed to be a Lorentz
connection, i.e., metricity holds,
\begin{equation}
D_c \eta_{ab}=0\label{eq:9},
\end{equation}
so that
$\Gamma_{abc}=\eta_{ad}{\Gamma^d}_{bc}=-\Gamma_{bac}$ and
\begin{equation}
  \label{eq:37}
  \Gamma_{abc}=K_{abc}+r_{abc},
\end{equation}
where
\begin{equation}
  \label{eq:99}
K_{abc}=\half(T_{bac}+T_{cab}-T_{abc})=-K_{bac}
\end{equation}
is the contorsion tensor, and
\begin{equation}
  r_{abc}=\half(D_{bac}+D_{cab}-D_{abc})=-r_{bac} \label{eq:96}
\end{equation}
the torsion-free Levi-Civita connection.
Furthermore,
\begin{equation}
  \label{eq:29}
  R_{abcd}=-R_{bacd}, \quad S_{ab}=0
\end{equation}
and
\begin{multline}
  \label{eq:30}
  R_{abcd}-R_{cdab}=\frac{3}{2}(D_{[b}T_{|a|cd]}+T_{af[b}{T^f}_{cd]}
-D_{[a}T_{|b|cd]}-T_{bf[a}{T^f}_{cd]}\\-D_{[d}T_{|c|ab]}-T_{cf[d}{T^f}_{ab]}
+D_{[c}T_{|d|ab]}+T_{df[c}{T^f}_{ab]}).
\end{multline}
The curvature scalar is defined by
${R}=\eta^{ab}{R}_{ab}$, the Einstein tensor by
\begin{equation}
  \label{eq:33}
  G_{ab}={R}_{(ab)}-\half \eta_{ab} {R}.
\end{equation}
Contracting \eqref{eq:28a} with $\eta^{bf}$ gives the contracted
Bianchi identities
\begin{equation}
  \label{eq:34b}
D_b {{G}^b}_{a}=\half
  {R^{bc}}_{da}{T^{d}}_{bc}+{{R}^b}_c{T^c}_{ab}  -\half D^b(D_c {T^c}_{ab}
+2D_{[a} {T^c}_{b]c}+{T^c}_{dc}{T^d}_{ab}) .
\end{equation}

\subsection{Variational principle for Einstein gravity}

The inclusion of torsion in the previous considerations allows one to
formulate a convenient action principle with Euler-Lagrange equations
that impose vanishing of torsion together with all NP equations. The
action is first order and of Cartan type. It involves as dynamical
variables $\phi^i$ the vielbein components ${e_a}^\mu$, the Lorentz
connection components in the non-holonomic frame $\Gamma_{abc}$,
together with a suitable set of auxiliary fields
${\mathbf R}_{abcd}=\mathbf{R}_{[ab][cd]},
\lambda^{abcd}=\lambda^{[ab][cd]}$,
\begin{equation}
S[\Gamma_{abc},{e_a}^\mu, {\mathbf R}_{abcd}, \lambda^{abcd} ]=k\int
\mathcal L,\label{eq:1}
\end{equation}
with $k^{-1}=-16\pi G$, where the minus sign is required for the
$(+--~-)$ convention adopted here and
\begin{equation}
\mathcal L=\star L,\quad  L=\mathbf{R}_{abcd}(\eta^{ac}\eta^{bd} - \lambda^{abcd})
   + \lambda^{abcd} R_{abcd}+2\Lambda^C,
\label{action NP}
\end{equation}
where $R_{abcd}=\eta_{ae}{R^e}_{bcd}$ is explicitly given in
\eqref{eq:12} as a function of the variables
${e_a}^\mu,\Gamma_{abc}$ and their first order derivatives, and
$\Lambda^C$ denotes the cosmological constant. For simplicity, we
put $k=1$ for the remainder of
this section.

The equations of motion for the auxiliary fields follow from equating
to zero the Euler-Lagrange derivatives of the $n$-forms
\begin{equation}
  \label{eq:11a}
  \begin{split}
   \frac{\delta \mathcal L}{\delta \mathbf R_{abcd}} & =
  -\star \left[\lambda^{abcd}-\half (\eta^{ac}\eta^{bd}-\eta^{ad}\eta^{bc})
    \right],\\
 \frac{\delta \mathcal L}{\delta {\mathbf
     \lambda^{abcd}}} & =-\star \left[\mathbf
   R_{abcd}-R_{abcd}\right].
  \end{split}
\end{equation}
They thus fix the auxiliary $\lambda$ fields in terms of the Minkowski
metric,
\begin{equation}
\lambda^{abcd}= \half
(\eta^{ac}\eta^{bd}-\eta^{ad}\eta^{bc})\equiv\lambda^{abcd}_\eta,\label{eq:119}
\end{equation}
and impose the definition of the Riemann tensor in terms of vielbein
and connection components as on-shell relations,
$\mathbf R_{abcd}=R_{abcd}$, which is desirable from the viewpoint of
the NP formalism. They can be eliminated by solving inside the
action. The resulting reduced action coincides with the standard
action associated to the Cartan formalism, up to an invertible change
of variables. More explicitly,
$L_C[{e_a}^\mu,\Gamma_{ab\nu}]=(R_{ab\mu\nu}{e_c}^\mu{e_d}^\nu\eta^{ac}
\eta^{bd}+2\Lambda^C)$. If we denote by a prime a function on which we
perfom the invertible change of variables
$\Gamma_{ab\mu}=\Gamma_{abc}{e^c}_\mu$, the reduced action is
$\bar S[{e_a}^\mu,\Gamma_{abc}]=\int \mathcal L'_C$ with
\begin{equation}
\mathcal L'_C=\star L'_C,\quad L'_C=(R_{abcd}\eta^{ac}\eta^{bd}+2\Lambda^C). \label{eq:18}
\end{equation}

The next equations of motion follow from the vanishing of
\begin{equation}
  \frac{\delta \mathcal L}{\delta \Gamma_{abc}}=2
  \star \left[D_f\lambda^{abcf}+\lambda^{abdf} ({T^h}_{fh}\delta^c_d+\half
    {T^c}_{df})\right]. \label{2}
\end{equation}
When putting $\lambda^{abcd}$ on-shell, they are equivalent to
vanishing of torsion, ${T^a}_{bc}=0$. It follows that
$\Gamma_{abc}=r_{abc}$, or equivalently that
${\Gamma^a}_{bc} ={e^a}_\nu {e_c}^\mu\nabla_\mu {e_b}^\nu$, where
$\nabla_\mu$ denotes the Christoffel connection. In other words, the
connection components are also auxiliary fields that can be expressed
in terms of vielbein components and eliminated by their own equations
of motion.

The last equations of motion follow from the vanishing of
\begin{equation}
  \frac{\delta \mathcal L}{\delta
    {e_a}^\mu}={e^b}_\mu\left[2\star(\lambda^{cdfa}R_{cdfb})-\frac{\delta
      \mathcal L}{\delta \Gamma_{cda}}\Gamma_{cdb}\right]
  - {e^a}_\mu\left[\star (\mathbf R +2\Lambda^C)+
    \lambda^{bcdf}\frac{\delta \mathcal L}{\delta\lambda^{bcdf}}\right].
\end{equation}
On-shell for the auxiliary fields, we have
\begin{equation}
  \label{eq:14}
  \frac{\delta \mathcal L}{\delta
    {e_a}^\mu}|_{\rm aux\ on-shell}=2\star {e^b}_\mu ({G^a}_b-\Lambda^C\delta^a_b),
\end{equation}
which imply the standard Einstein equations.

Finally, let us also note that the equations of motion in the
Cartan formalism with spin coefficients $\Gamma_{abc}$ as variables
are determined by
\begin{equation}
  \label{eq:26}
  \begin{split}
  \vddl{\mathcal L'_C}{\Gamma_{abc}} &=-\star
  ({T^{ha}}_h\eta^{bc}-{T^{hb}}_h\eta^{ac}-{T^{cab}}), \\
  \vddl{\mathcal L'_C}{{e_a}^\mu} &={e^b}_\mu[2\star
  {R^a}_b-\vddl{\mathcal L'_C}{\Gamma_{cda}}\Gamma_{cdb}]-{e^a}_\mu\star (
  R+2\Lambda^C),
\end{split}
\end{equation}
and that they are explicitly related to the standard ones through
\begin{equation}
  \label{eq:19}
\vddl{\mathcal L'_C}{\Gamma_{abc}}={e^c}_\mu(\vddl{\mathcal L_C}{\Gamma_{ab\mu}})',\quad
\vddl{\mathcal L'_C}{{e_{a}}^\mu}=(\vddl{\mathcal L_C}{{e_{a}}^\mu})'
-\Gamma_{bcd}{e^d}_\mu{e^a}_\nu(\vddl{\mathcal L_C}{\Gamma_{bc\nu}})'.
\end{equation}

\subsection{Relation to Newman-Penrose formalism in 3 dimensions}
\label{sec:generalities}

The vacuum Einstein equations $G_{ab}-\Lambda^C\eta_{ab}\approx 0$ imply
that
\begin{equation}
  \label{eq:80}
  R\approx -\frac{2d}{d-2}\Lambda^C,\quad R_{(ab)}\approx
  -\frac{2}{d-2}\Lambda^C\eta_{ab}.
\end{equation}

In three dimensions, in the absence of torsion (where in particular
$R_{(ab)}=R_{ab}$), one may show that
\begin{equation}
  \label{eq:79}
  R_{abcd}=\eta_{ac}R_{bd}-\eta_{ad}R_{bc}-\eta_{bc}R_{ad}+\eta_{bd}R_{ac}-\half
  (\eta_{ac}\eta_{bd}-\eta_{ad}\eta_{bc})R,
\end{equation}
so that, \eqref{eq:80} implies
\begin{equation}
  \label{eq:84}
  R_{abcd}\approx -\Lambda^C(\eta_{ac}\eta_{bd}-\eta_{ad}\eta_{bc}).
\end{equation}
In applications, we choose \bea \eta_{ab}=\begin{pmatrix} 0&1&0
  \\ 1&0&0 \\ 0&0&-\half \end{pmatrix}, \eea
as in a version of the Newman-Penrose formalism adapted to three
dimension \cite{Milson:2012ry,Barnich:2016lyg}.

\subsection{Relation to Newman-Penrose formalism in 4 dimensions}
\label{sec:relat-newm-penr}

In four spacetime dimensions, the tetrads
$e_1=l,\,e_2=n,\,e_3=m,\,e_4=\bar m$ in the NP formalism are choosen
as null vectors, $e_a\cdot e_b=\eta_{ab}$ with
\begin{equation}
\eta_{ab}=\eta^{ab} = \begin{pmatrix}
0 & 1 & 0 & 0 \\
1 & 0 & 0 & 0 \\
0 & 0 & 0 &-1 \\
0 & 0 & -1 &0 \\
\end{pmatrix}.
\end{equation}
The components of the Lorentz connection are traded for the spin
coefficients,
\begin{equation}
  \label{spinconnection}
  \begin{split}
    &\kappa=\Gamma_{311},\;\;\pi=-\Gamma_{421},\;\;\epsilon=\half(\Gamma_{211} - \Gamma_{431}),\\
    &\tau=\Gamma_{312},\;\;\nu=-\Gamma_{422},\;\;\gamma=\half(\Gamma_{212} - \Gamma_{432}),\\
    &\sigma=\Gamma_{313},\;\;\mu=-\Gamma_{423},\;\;\beta=\half(\Gamma_{213} -\Gamma_{433}),\\
    &\rho=\Gamma_{314},\;\;\lambda=-\Gamma_{424},\;\;\alpha=\half(\Gamma_{214}
    - \Gamma_{434}).
\end{split}
\end{equation}
The other half of the spin coefficients are denoted with a bar on the
symbols of the left hand sides and obtained by exchanging the index $3$
and $4$ on the right hand sides. The Weyl tensor $C_{abcd}$ is encoded
in terms of
\begin{equation}
  \label{weyl}
\Psi_0=-C_{1313},\;\;\Psi_1=-C_{1213},\;\;\Psi_2=-C_{1342},\;\;\Psi_3=-C_{1242},\;\;\Psi_4=-C_{2324},
\end{equation}
with the same rule as above for $\bar\Psi_i$, $i=0,\dots, 4$, while the
Ricci tensor is organized as
\begin{equation}
\begin{array}{lll}
\Phi_{00}=-\half R_{11}, & \Phi_{11}=-\dfrac{1}{4}(R_{12}+R_{34}), & \Phi_{22}=-\half R_{22},\\
\Phi_{02}=-\half R_{33}, & \Phi_{01}=-\half R_{13}, & \Phi_{12}=-\half R_{23},\\
\Phi_{20}=-\half R_{44}, & \Phi_{10}=-\half
                           R_{14},& \Phi_{21}=-\half R_{24},\\
  & \Lambda=\dfrac{1}{24}R=\dfrac{1}{12} (R_{12}-R_{34}). &
  \end{array}
\end{equation}
There is no torsion in the NP approach, ${T^a}_{bc}=0$. In this case,
the vacuum Einstein equations in flat space are equivalent to the
vanishing of the $\Phi$'s. The equations governing the NP quantities
can then be interpreted as follows. (i) The metric equations express
commutators of tetrads in terms of spin coefficients. This is the
first of \eqref{eq:13} when taking into account that
${D^a}_{bc}=2{\Gamma^a}_{[cb]}$ in the absence of torsion. (ii) The
spin coefficient equations express directional derivatives of spin
coefficicents in terms of spin coefficients and the Weyl and Ricci
tensors. In the torsion-free case, they are equivalent to the
definition of $R_{abcd}$ in the second of \eqref{eq:12}. (iii) The
Bianchi identities express directional derivatives of the $\Psi's$ and
$\Phi$'s in terms of spin coefficents and $\Psi$'s and $\Phi$'s. They
are equivalent to the second of \eqref{eq:24} in the absence of
torsion\footnote{The parametrization of class III rotations after
  equation (6.9) of \cite{Barnich:2016lyg} should be corrected to
  $l'=e^{E_R}l, n'=e^{-E_R}n$.}.

\subsection{Improved gauge transformations and Noether identities}

Diffeomorphisms and local Lorentz transformations are extended in
a natural way to the auxiliary fields. If ${\xi'}^\mu,{{\omega'}^{a}}_b=-{\omega'_b}^a$
denote parameters for the infinitesimal transformations, they act on
the fields as
\begin{equation}
  \label{eq:15}
  \begin{split}
    \delta_{\xi',\omega'}{e_a}^\mu &={\xi'}^\nu\partial_\nu
    {e_a}^\mu-\d_\nu{\xi'}^\mu{e_a}^\nu +{\omega'_a}^b{e_b}^\mu,\\
\delta_{\xi', \omega'} \Gamma_{a b c} &= {\xi'}^\nu \partial_\nu \Gamma_{a
  b c} - D_c {\omega'}_{a b} + {\omega'_c}^{d}\Gamma_{abd} ,\\
\delta_{\xi', \omega'} \mathbf R_{abcd} &={\xi'}^\nu \partial_\nu \mathbf R_{abcd}
+ {\omega'}\indices{_a^f}\mathbf R_{fbcd} + {\omega'}\indices{_b^f}\mathbf R_{afcd}
+ {\omega'}\indices{_c^f}\mathbf R_{abfd} + {\omega'}\indices{_d^f} \mathbf R_{abcf}, \\
\delta_{\xi', \omega'} \lambda^{abcd} &={\xi'}^\nu \partial_\nu
\lambda^{abcd}
+ {\omega'}^a_{\;\;f} \lambda^{fbcd}  + {\omega'}^b_{\;\;f} \lambda^{afcd}
+ {\omega'}^c_{\;\;f} \lambda^{abfd}  + {\omega'}^d_{\;\;f}\lambda^{abcf}.
\end{split}
\end{equation}
In terms of the redefined gauge parameters, which are now spacetime
scalars, and thus in agreement with the general strategy of the NP
approach,
\begin{equation}
\xi^a={e^a}_\mu{\xi'}^\mu,\quad
\omega\indices{_a^b}={\omega'_a}^b+{\xi'}^\mu{\Gamma^b}_{ac}{e^c}_\mu,\label{eq:70}
\end{equation}
these gauge transformations become
\begin{equation}
  \label{eq:23}
  \begin{split}
    \delta_{\xi,\omega}{e_a}^\mu &=(\xi^c {T^b}_{ac}
    -D_a\xi^b+{\omega}\indices{_a^b}){e_b}^\mu,\\
    \delta_{\xi,\omega} \Gamma_{a b c} &= -\xi^d
    R_{abcd}+ (\xi^f{T^d}_{cf}-D_c
    \xi^d+\omega\indices{_c^d})\Gamma_{abd}
    -D_c\omega_{ab},\\
    \delta_{\xi,\omega} \mathbf R_{abcd} &= \xi^f D_f \mathbf R_{abcd} +
    {\omega}\indices{_a^f}\mathbf R_{fbcd} +
    \omega\indices{_b^f}\mathbf R_{afcd}
    + \omega\indices{_c^f}\mathbf R_{abfd} +
    \omega\indices{_d^f} \mathbf R_{abcf} , \\
    \delta_{\xi,\omega} \lambda^{abcd} &=\xi^f D_f
    \lambda^{abcd} + \omega\indices{^a_f} \lambda^{fbcd} +
    \omega\indices{^b_f} \lambda^{afcd} + \omega\indices{^c_f} \lambda^{abfd}
    + \omega\indices{^d_f}\lambda^{abcf}.
\end{split}
\end{equation}
We refer to the latter as ``improved gauge
  transformations'' since they involve the derivatives of the objects
  that are being transformed only in the form of tensors.

Isolating the undifferentiated gauge parameters by dropping the
exterior derivative of an $n-1$ form, the invariance of action
\eqref{eq:1} under these transformations leads to the Noether
identities. Since the change of gauge parameters is invertible, the
identities associated to both sets are equivalent. We can thus
concentrate on this second set. For later use, note that
\begin{equation}
\delta_{\xi,\omega}\Gamma_{abc}-(\delta_{\xi,\omega}{e_c}^\mu)
{e^d}_\mu\Gamma_{abd}=-\xi^d
R_{abcd}-D_c\omega_{ab}.\label{eq:91}
\end{equation}

When using \eqref{eq:46}, the Noether identities associated to the
Lorentz parameters $\omega_{ab}$ become
\begin{multline}
  2 \frac{\delta \mathcal L}{\delta \mathbf R_{[a|cdf|}} {\mathbf
    R^{b]}}_{cdf} + 2 \frac{\delta \mathcal L}{\delta \mathbf R_{cd[a|f|}}
  {{\mathbf R_{cd}}^{b]}}_f + 2 \frac{\delta \mathcal L}{\delta
    \lambda^{fhcd}} \eta^{f[a}\lambda^{b]hcd}
  + 2 \frac{\delta \mathcal L}{\delta \lambda^{cdfh}} \eta^{f[a}\lambda^{|cd|b]h} \\
  +\frac{\delta \mathcal L}{\delta {e_{[a}}^\mu} e^{b]\mu} + \frac{\delta
    \mathcal L}{\delta \Gamma_{cd[a}} {\Gamma_{cd}}^{b]} + \star
  \big[(D_c +{T^c}_{cf})(\star^{-1} \frac{\delta \mathcal L}{\delta \Gamma_{abc}})\big]
  =0. \label{eq:NL}
\end{multline}
while the Noether identities for the vector fields $\xi^f$ read
\begin{multline}
  \frac{\delta \mathcal L}{\delta \mathbf R_{abcd}} D_f \mathbf R_{abcd}
  + \frac{\delta \mathcal L}{\delta \lambda^{abcd}} D_f \lambda^{abcd} +
  \frac{\delta \mathcal L}{\delta {e_a}^\mu} {T^b}_{af}{e_b}^\mu
  +\frac{\delta \mathcal L}{\delta \Gamma_{abc}}({T^d}_{cf}\Gamma_{abd} -R_{abcf})\\
  + \star \big[(D_c+{T^h}_{ch})\star^{-1}( \frac{\delta \mathcal
    L}{\delta {e_c}^{\mu}}{e_f}^\mu +\frac{\delta \mathcal
    L}{\delta \Gamma_{abc}}\Gamma_{abf})\big] =0.\label{eq:ND}
\end{multline}

It follows from general results on auxiliary fields that these Noether
identities are equivalent to those of the standard Cartan formulation,
which have been investigated and related to the Bianchi identities in
\cite{Barnich:2016rwk}. More explicitly, we have $L=L'_C+A$ with
$A=[(\mathbf
R_{abcd}-R_{abcd})(\eta^{ac}\eta^{bd}-\lambda^{abcd})]$. Identity
\eqref{eq:NL} for $L$ replaced by $A$ is equivalent to
\eqref{eq:27}. This then implies that \eqref{eq:NL} reduces to
\begin{equation}
  \label{eq:16}
  \frac{\delta \mathcal L'_C}{\delta {e_{[a}}^\mu} e^{b]\mu}
+ \frac{\delta \mathcal L'_C}{\delta \Gamma_{cd[a}} {\Gamma_{cd}}^{b]}
+ \star \big[(D_c+{T^f}_{cf}) (\star^{-1} \frac{\delta \mathcal
  L'_C}{\delta \Gamma_{abc}})\big] =0,
\end{equation}
which in turn is also equivalent to \eqref{eq:27}.

Identity \eqref{eq:ND} for $L$ replaced by $A$ is equivalent to the
second of \eqref{eq:24}. This then implies that \eqref{eq:ND} reduces
to
\begin{multline}
  \label{eq:28}
 \frac{\delta \mathcal L'_C}{\delta {e_a}^\mu} {T^b}_{af}{e_b}^\mu
  +\frac{\delta \mathcal L'_C}{\delta \Gamma_{abc}}({T^d}_{fc}\Gamma_{abd} -R_{abcf})
  \\+ \star\big[ (D_c+{T^h}_{ch})\star^{-1} (\frac{\delta \mathcal L'_C}{\delta {e_c}^{\mu}}{e_f}^\mu
  +\frac{\delta \mathcal L'_C}{\delta \Gamma_{abc}}\Gamma_{abf})\big] =0,
\end{multline}
which in turn is equivalent to \eqref{eq:34b}.

\subsection{Closed co-dimension 2 forms}
\label{sec:appl-non-holon}

\paragraph{Presymplectic 1 form potential}

The presymplectic 1 form potential associated to the action
\eqref{action NP} is given by
\begin{equation}
  \label{eq:93}
  a=2{\mathbf e}\lambda^{abcd}{e_c}^\mu d_V
  \Gamma_{ab\nu}{e^\nu}_d{e_c}^\mu\star (dx_\mu),
\end{equation}
where $d_V
  \Gamma_{ab\nu}{e^\nu}_d=d_V\Gamma_{abd}-\Gamma_{abf}{e^f}_\nu d_V
  {e_d}^\nu$.

\paragraph{Weakly vanishing Noether current}

In the non-holonomic version of the Cartan formulation with auxiliary
fields, we have
$\phi^i=(\mathbf R_{abcd},\lambda^{abcd},\Gamma_{abc},{e_a}^\mu)$ and
$f^\alpha=(\omega_{ab},\xi^a)$, while the weakly vanishing
Noether currents are encoded in the $1$-form
\begin{equation}
  \label{eq:35}
 \begin{split}
  S_{\omega,\xi} & =-
  \star^{-1}\big[ \vddl{\mathcal
    L}{\Gamma_{abc}}(\omega_{ab}+ \Gamma_{abf}\xi^f)+\vddl{\mathcal
    L}{{e_{c}}^\mu}{e_f}^\mu\xi^f\big]{}^*e_c\\
  & =4\delta^R K_{\omega} +S^R_{\xi},\
\end{split}
\end{equation}
where
\begin{equation}
  \label{eq:45}
  \begin{split}
    K_{\omega}&=-\frac{1}{4} \omega_{ab}{\lambda^{ab}}_{cd}{}^*e^{c}{}^*e^{d},\\
    S^R_{\xi}&=-2\big[\lambda^{abcd}(R_{abcf}\delta^h_d-\half
R_{abcd}\delta^h_f)+\Lambda^C\delta^h_f\\
  & \hspace{5cm} +\half(\lambda^{abcd}-\lambda^{abcd}_\eta)\mathbf
  R_{abcd}\delta^h_f\big] \xi^f{}^*e_h.
\end{split}
\end{equation}

\paragraph{Co-dimension 2 form}

Using \eqref{eq:34} and \eqref{eq:48} in form degree $n-2$ gives
\begin{equation}
I^{n-1}_\varphi[\star (\delta^R K_{\omega(x)})]=\delta^V (\star
K_{\omega(x)})
-d\big(I_\varphi^{n-2}[(\star
K_{\omega(x)})]\big),\label{eq:53}
\end{equation}
where $I_\varphi^{n-2}[(\star
K_{\omega(x)})]=0$ since
$K_{\omega(x)}$ involves no derivatives. When using \eqref{eq:63},
we also have
\begin{equation}
  \label{eq:61}
 I^{n-1}_\varphi[\star
 S^R_{\xi(x)}]=\star\Big[(\delta\Gamma_{ab\mu}){e_c}^\mu\big({\lambda^{ab}}_{df}\xi^c(x) {}^*e^{d}
 {}^*e^{f}+2{\lambda^{abc}}_d\xi_f(x){}^*e^{d} {}^*e^{f}\big)\Big],
\end{equation}
with the understanding that
\begin{equation}
  \label{eq:128}
  \delta \Gamma_{ab\mu}{e_c}^\mu=\delta\Gamma_{abc}-\Gamma_{abd}{e^d}_\mu\delta e^\mu_c,
\end{equation}
in terms of the fundamental variables
$\Gamma_{abc},{e_d}^\mu$ used here.

When putting everything together, we get
\begin{multline}
  \label{eq:59}
  \star k_{\omega,\xi}
  =-\omega_{ab}\delta^V(\frac{1}{(n-2)!}\lambda^{abcd}\epsilon_{cdb_3\dots
  b_n}{}^*e^{b_3}\dots {}^*e^{b_n})\\
+\frac{1}{(n-2)!}(
\lambda^{abcd}\xi^f+2\lambda^{abf[c}\xi^{d]})\delta\Gamma_{ab\mu}{e_f}^\mu
\epsilon_{cdb_3\dots b_n} {}^*e^{b_3}\dots {}^*e^{b_n}.
\end{multline}
Inserting the first equation of motion \eqref{eq:119} gives
\begin{equation}
  \label{eq:59a}
  \star k_{\omega,\xi}
  =\frac{1}{(n-2)!}(-\omega^{ab}\delta^V
+\delta{\Gamma^{ab}}_\mu{e_f}^\mu\xi^f+2\delta{\Gamma^{f[a}}_\mu{e_f}^\mu\xi^{b]})
\epsilon_{abc_3\dots c_n} {}^*e^{c_3}\dots {}^*e^{c_n}.
\end{equation}
When taking into account the redefinitions in \eqref{eq:70}, together
with \eqref{eq:128}, one finds the same expression as the one obtained
in the standard Cartan formalism in (3.49) of
\cite{Barnich:2016rwk}.
An equivalent form is
\begin{empheq}[box=\fbox]{equation}
  \label{eq:130}
  \begin{split}
  & \star k_{\xi,\omega}=-\delta
  K^K_{\omega}+K^K_{\delta\omega}+
  K^K_{\delta\Gamma_{\rho}\xi^\rho}
  -(\xi^b\dover{}{{}^*e^{b}})\Theta,\\
  & \Theta=\star(2\delta{{\Gamma^{ac}}_\rho}{e_c}^\rho {}^*e_a),\quad
  K^K_{t}=\star (t_{ab}{}^*e^{a}{}^*e^{b}),
\end{split}
\end{empheq}
with
\begin{equation}
  \label{eq:125}
  K^K_{\delta\omega}+K^K_{\delta\Gamma_\mu\xi^\mu}-(\xi^b\dover{}{{}^*e^{b}})\Theta=\star\big[(\delta\omega_{ab}+
  \delta\Gamma_{ab\rho}\xi^\rho-2\delta\Gamma\indices{_{[a|}^c_{\mu}}
  {e_c}^{\mu}\xi_{|b]}){}^*e^{a}{}^*e^{b}\big].
\end{equation}

\paragraph{Breaking}

Using \eqref{eq:75}, the breaking is explicitly given by
\begin{multline}
  \label{eq:25}
  b_{\xi,\omega}=2  \big\{  \big[\delta_{\xi,\omega}\lambda^{abcd}
      +(2\lambda^{abf[d}{e^{c]}}_\mu+\lambda^{abdc}{e^f}_\mu)\delta_{\xi,\omega}
      {e_f}^\mu\big]\delta\Gamma_{ab\nu}{e_c}^\nu\\
-(\delta_{\xi,\omega}\longleftrightarrow \delta) \big\} 
      {}^*e_d.
\end{multline}
On-shell,
$\delta_{\xi,\omega}\lambda^{abcd}_\eta=0=\delta\lambda^{abcd}_\eta$, so that
the breaking reduces to
\begin{empheq}[box=\fbox]{multline}
  \label{eq:25a}
  b_{\xi,\omega}=2\big[\delta_{\xi,\omega}
      {e_b}^\mu\delta{\Gamma^{ab}}_\nu{e_a}^\nu{e^c}_\mu+\delta_{\xi,\omega}
      {e_a}^\mu\delta{\Gamma^{ac}}_\mu-\delta_{\xi,\omega}\ln {\mathbf
        e}\,\delta{\Gamma^{cb}}_\nu{e_b}^\nu
     \\ -(\delta_{\xi,\omega}\longleftrightarrow\delta)\big]{}^*e_c.
  \end{empheq}

Note that the RHS of \eqref{eq:59a} and \eqref{eq:25a} need to
be multiplied by $k=-(16\pi G)^{-1}$.

\paragraph{Exact reducibility parameters}

General considerations on (generalized) auxiliary fields imply that,
on-shell, reducibility parameters should be given by Killing vectors
$\bar \xi^a$ of the metric. Let us see how this comes about here.

Merely the first of \eqref{eq:23} encodes gauge
transformations of fields that are not auxiliary. The associated
equation
$\delta_{\bar\omega,\bar\xi}{e_a}^\mu\approx 0$ is equivalent to
\begin{equation}
  \label{eq:8}
  D_{(a}\bar\xi_{b)}-\bar\xi^cT_{(ba)c}\approx 0,\quad
  \bar\omega_{ab}\approx D_{[a}\bar\xi_{b]}-\bar\xi^cT_{[ba]c}.
\end{equation}
On-shell when torsion vanishes, the first indeed requires $\bar\xi^a$
to be a Killing vector, while the second uniquely fixes the Lorentz
parameters in terms of it. In particular,
\begin{equation}
\bar\omega_{ab}\approx
D_a\bar\xi_b\approx -D_b\bar\xi_a. \label{eq:76}
\end{equation}

The other equations impose no additional constraints. Indeed,
$\delta_{\bar\omega,\bar\xi}\lambda^{abcd}\approx 0$ is satisfied
identically on account of the skew-symmetry of
$\bar\omega^{ab}$. Instead of
$\delta_{\bar\xi,\bar\omega}\Gamma_{abc}\approx 0$ we can consider the
combination \eqref{eq:91}. Requiring this to vanish on-shell amounts
to
\begin{equation}
  \label{eq:78}
  D_c\omega_{ab}\approx -\bar\xi^d R_{abcd},
\end{equation}
which holds as a consequence of the second equation in \eqref{eq:8},
when using that
\begin{equation}
D_aD_b\bar\xi_c\approx
{R^d}_{abc}\bar\xi_d\label{eq:77},
\end{equation}
which can be shown as in \cite{Wald:1984rg} Appendix C.3, and when
using also \eqref{eq:30}. Finally,
$\delta_{\bar\xi,\bar\omega}{\mathbf R}_{abcd}\approx 0$, reduces
on-shell to
\begin{equation}
  \label{eq:31}
  \bar\xi^f D_f  R_{abcd} +
    {\bar\omega}\indices{_a^f} R_{fbcd} +
    \bar\omega\indices{_b^f} R_{afcd}
    + \bar\omega\indices{_c^f} R_{abfd} +
    \bar\omega\indices{_d^f}  R_{abcf}\approx 0.
\end{equation}
This equation holds because one can show that, on-shell, the left hand
side is equal to its opposite when using the previous relations
\eqref{eq:76}, \eqref{eq:77} together with the Bianchi identities
\eqref{eq:24} and the on-shell symmetries of the Riemann tensor.

\section{Application to asymptotically flat 4d gravity}

\subsection{Solution space}
\label{sec:solution-space-1}

Four-dimensional asymptotically flat spacetimes at null infinity in
the NP formalism have been studied in
\cite{Newman:1961qr,Newman:1962cia,Exton1969} (see
\cite{Barnich:2016lyg} for a summary and conventions appropiate to the
current context). One uses standard coordinates $x^\mu = (u, r, x^A)$,
$x^A = (\zeta, \bar{\zeta})$,  where $u$ labels null
  hypersurfaces, $r$ is an affine parameter along the generating null
  geodesics and $x^A$ are stereographic coordinates in the simplest
  case when future null infinity is a sphere.  In the notations of
section \ref{sec:relat-newm-penr}, the Newman-Unti solution space is
entirely determined by the conditions
\begin{equation}
\begin{split}
  &\kappa = \epsilon = \pi = 0,\quad \rho=\xbar\rho, \quad \tau=\bar\alpha+\beta,\\
  &l = \frac{\partial}{\partial r}, \quad n = \frac{\partial}{\partial
    u} + U \frac{\partial}{\partial r} + X^A \frac{\partial}{\partial
    x^A} , \quad m = \omega \frac{\partial}{\partial r} + L^A
  \frac{\partial}{\partial x^A},
\end{split}
\label{gauge conditions}
\end{equation}
where $U$, $X^A$, $\omega$ and $L^A$ are arbitrary functions of the
coordinates, together with the fall-off conditions
\begin{equation}
\begin{split}
&X^A = \cO(r^{-1}), \quad \Psi_0 = \Psi^0_0 r^{-5} +
\mathcal{O}(r^{-6}), \quad \rho = -\dfrac{1}{r}+\cO(r^{-3}) , \quad \tau = \cO (r^{-2}),\\
&g_{AB} dx^A dx^B = -2 r^2 \frac{d \zeta d\bar{\zeta} }{P \bar{P}} +
\mathcal{O}(r).
\end{split}
\label{fall-off conditions}
\end{equation}
Here $\Psi^0_0$ and $P$ are arbitrary complex functions of
$(u,\zeta,\bar{\zeta})$. Below we will also use the real function
$\varphi(u,\zeta,\bar\zeta)$ defined by $P\bP=2 e^{-2\varphi}$. The
associated asymptotic expansion of the solution space in terms of
$\Psi_0 (u_0, r, \zeta, \bar{\zeta})$ ,
$(\Psi^0_2 + \bar{\Psi}^0_2)(u_0, \zeta, \bar{\zeta})$,
$\Psi^0_1 (u_0, \zeta, \bar{\zeta})$ at fixed $u_0$ and of the
asymptotic shear $\sigma^0(u, \zeta, \bar{\zeta})$ and the conformal
factor $P(u, \zeta, \bar{\zeta})$ is summarized in Appendix \ref{NP
  solution}. This data characterizing solution space is collectively
denoted by $\chi$.

On a space-like cut of $\mathscr{I}^+$, we
use coordinates $\zeta, \bar{\zeta}$, and the (rescaled) induced
metric
\begin{equation}
d\bar{s}^2 = -\bar{\gamma}_{AB} dx^A dx^B =  -2 (P \bar{P})^{-1} d\zeta d\bar{\zeta},
\end{equation} with $\bar{P}P> 0$. For the unit sphere, we have $\zeta
= \cot \frac{\theta}{2} e^{i\phi}$ in terms of standard spherical
coordinates and
\begin{equation}
P_S (\zeta, \bar{\zeta}) = \frac{1}{\sqrt{2}}(1+ \zeta \bar{\zeta}).\label{unit}
\end{equation} The covariant derivative on the 2 surface is
encoded in the operators
\begin{equation}\begin{split}
    &\eth \eta^s=P\bP^{-s}\bp(\bP^s \eta^s)=P\bp \eta^s
    + s P \bp \ln \bP \eta^s=P\bp \eta^s + 2 s\xbar\alpha^0 \eta^s,\\
    &\xbar\eth \eta^s=\bP P^{s}\p(P^{-s} \eta^s)=\bP\p \eta^s
    - s \bP \p \ln P \eta^s=\bP\p \eta^s -2 s \alpha^0 \eta^s,
\end{split}\end{equation}
where $s$ is the spin weight of the field $\eta$ and
$\p=\p_{\zeta},\bar\p=\p_{\bar\zeta}$. The spin and conformal weights
of relevant fields (see Appendix \ref{NP
  solution} and section \ref{sec:acti-symm-solut-1}) are listed in Table \ref{t1}.
\begin{table}[h]
\caption{Spin and conformal weights}\label{t1}
\begin{center}\begin{tabular}{c|c|c|c|c|c|c|c|c|c|c|c|c|c|c|c|c|c}
& $\eth$ & $\p_u$ & $\gamma^0$ & $\nu^0$ & $\mu^0$ & $\sigma^0$ &
   $\lambda^0$
  & $\Psi^0_4$ &  $\Psi^0_3$ & $\Psi^0_2$ & $\Psi^0_1$ & $\Psi_0^0$ & $\cY$   \\
\hline
s & $1$& $0$& $0$& $-1$& $0$& $2$& $-2$  &
  $-2 $&  $-1$ & $0$ & $1$ & $2$ & $-1$  \\
\hline
w  & $-1$& $-1$& $-1$& $-2$& $-2$& $-1$& $-2$  &
  $-3 $&  $-3$ & $-3$ & $-3$ & $-3$ & $1$  \\
\end{tabular}\end{center}\end{table}
Complex conjugation transforms the spin weight into its opposite and
leaves the conformal weight unchanged. The operators $\eth$,
$\overline{\eth}$ raise respectively lower the spin weight by one
unit. The Laplacian is
$\bar{\Delta} = 4 e^{-2 \varphi} \partial \bar{\partial} = 2 \xbar\eth
\eth$. Note that $P$ is of spin weight $1$ and ``holomorphic'',
$\xbar\eth P = 0$ and that
\begin{equation}
[\xbar\eth, \eth] \eta^s = \frac{s}{2} R \eta^s,
\end{equation} with $R= -4\mu^0=\bar{\Delta} \ln (P\bar{P})$, $R_S
=2$. We also have
\begin{equation}
[\partial_u, \eth] \eta^s =-2 (\xbar\gamma^0 \eth + s \eth
\gamma^0)\eta^s,\quad [\p_u,\xbar\eth]\eta^s=-2(\gamma^0\xbar\eth -
s\xbar\eth\xbar\gamma^0)\eta^s.
\end{equation}

The components of the inverse metric associated to the tetrad given in
\eqref{gauge conditions} is
\begin{equation*}
g^{0\mu} = \delta^\mu_1,~ g^{rr} = 2(U - \omega \overline{\omega}),~
g^{rA} = X^A - (\overline{\omega} L^A + \omega \overline{L}^A),~
g^{AB} = - (L^A \overline{L}^B + L^B \overline{L}^A ).
\end{equation*} Note furthermore that if $L_A = g_{AB} L^B$ with
$g_{AB}$ the two dimensional metric inverse to $g^{AB}$, then $L^A
\overline{L}_A = -1$, $L^A L_A = 0 = \overline{L}^A
\overline{L}_A$. The co-tetrad is given by
\begin{equation}
\begin{split}
{}^*e^{1} = - [U + X^A (\omega \overline{L}_A + \overline{\omega}
L_A)]du + dr + (\omega \overline{L}_A + \overline{\omega} L_A) dx^A, &
\\
{}^*e^{2} = du, \quad {}^*e^{3} = X^A \overline{L}_A du -
\overline{L}_A dx^A, \quad {}^*e^{4} = X^A L_A du - L_A dx^A. &
\end{split}
\label{cotetrad 4d case}
\end{equation}

\subsection{Residual gauge transformations}
\label{sec:resid-infn-gauge}

The parameters of residual gauge transformations that preserve the
solution space are entirely determined by asking that conditions
\eqref{gauge conditions} and \eqref{fall-off conditions} be preserved
on-shell. This is worked out in detail in Appendix \ref{ASG} where it
is shown that these parameters are given by
\begin{equation}
  f(u,\zeta,\bar\zeta),\quad Y^\zeta=Y(\zeta),\quad
  Y^{\bar\zeta}=\bar Y(\bar\zeta),\quad
  \omega'^{34}_0(u,\zeta,\bar\zeta).
  \label{orig}
\end{equation}
The associated residual gauge
transformations are explicitly determined by the gauge parameters,
\begin{equation}
\begin{split}
\xi'^u = f(u,\zeta,\bar{\zeta}),\quad
\xi'^A= Y^A(x^A) - \p_B f \int^{+\infty}_r dr[L^A \bL^B +  \bL^AL^B ], \\
\xi'^r=-\p_u f r + \frac{1}{2} \bar{\Delta}f - \p_A f
\int^{+\infty}_r dr[\omega \bL^A + \bomega L^A + X^A], \\
\end{split} \label{resiii1}
\end{equation}
and
\begin{equation}
\begin{split}
\omega'^{12}=& \p_u f + X^A \p_A f, \quad
\omega'^{23}= \bL^A \p_A f,\quad
\omega'^{24}= L^A \p_A f, \\
\omega'^{13}=&(\gamma^0 + \bar{\gamma}^0)\bar{P} {\partial} f -
\bar{P} \partial_u {\partial}f + \p_A f \int^{+\infty}_r dr[\lambda
L^A + \mu \bL^A], \\
\omega'^{14}=&(\gamma^0 + \bar{\gamma}^0)P \bar{\partial} f - P
\partial_u \bar{\partial}f + \p_A f \int^{+\infty}_r dr[\bar{\lambda}
\bL^A + \bar{\mu} L^A], \\
\omega'^{34}=&  \omega'^{34}_0(u,\zeta, \bar{\zeta}) - \partial_A f
\int^{+\infty}_{r} dr [(\bar{\alpha} - \beta ) \bar{L}^A +
(\bar{\beta} - \alpha) L^A].
\end{split} \label{resiii2}
\end{equation}
For the computations below, the leading orders of their asymptotic
on-shell expansions are also useful. When the solutions
  discussed in Appendix \ref{NP
  solution} are inserted, one obtains
\be\begin{split}\label{asymmetry1} \xi'^u=& f,
  \quad 
\xi'^\zeta= Y-\frac{\bP \eth f}{r}+\frac{\sigma^0\bP \xbar \eth
  f}{r^2}+O(r^{-3}),\quad \xi'^{\bar\zeta}=\overline{\xi'^\zeta},\\
\xi'^r=& -r\p_u f+ \half \bDelta f -\frac{\xbar\eth \sigma^0 \xbar\eth
  f + \eth \xbar\sigma^0 \eth f}{r}+O(r^{-2}),
\end{split}\ee
and
\be\begin{split}\label{asymmetry2}
\omega'^{12}=& \p_u f+O(r^{-3}),\quad
\omega'^{23}= \frac{\xbar \eth f}{r} - \frac{\xbar\sigma^0 \eth
  f}{r^2} + \frac{\sigma^0\xbar\sigma^0 \xbar\eth f}{r^3} +O(r^{-4}),
\\
\omega'^{13}=& (\gamma^0+\xbar\gamma^0) \xbar \eth f- \xbar\eth \p_u f
+ \frac{\lambda^0\eth f + \mu^0 \xbar\eth f}{r} \\ &\hspace{3cm}-
\frac{\xbar\sigma^0
  \mu^0 \eth f + \sigma^0\lambda^0\xbar\eth f}{r^2} - \frac{\Psi^0_2
  \xbar\eth f}{2r^2}+ O(r^{-3}),\\
\omega'^{34}=& \omega'^{34}_0 + \frac{\bar{P}
  \partial \ln P \eth f-P \bar{\partial} \ln \bar{P}
  \overline{\eth} f }{r} \\ &\hspace{3cm}+
\frac{P
  \bar{\partial} \ln \bar{P} \bar{\sigma}^0 \eth f
  -\bar{P} \partial \ln P \sigma^0 \overline{\eth} f }{r^2}  +  O(r^{-3}),
\end{split}\ee
with $\omega'^{24}=\overline{\omega'^{23}}$,
$\omega'^{14}=\overline{\omega'^{13}}$,
$\omega'^{34}=-\overline{\omega'^{34}}$.

\subsection{Residual symmetry algebra}
\label{sec:acti-symm-solut}

A direct application of the procedure outlined in section
\ref{sec:residual-symmetries} then gives the variation of the free
data parametrizing solution space under residual gauge transformation
in terms of the parametrization provided by \eqref{orig}. In
particular, one finds
\begin{equation}
-\delta_{f,Y,\omega'_0} P = P \partial_u f + f \partial_u P + Y \partial
P + \bar{Y}\bar{\partial} {P}- P \bar{\partial} \bar{Y} + P
\omega'^{34}_0,
\label{conformal factors}
\end{equation}
together with the variation of the rest of the free data and derived
quantities that is given in Appendix \ref{sec:oracti}.

In order to make these variations more transparent, it is useful to
re-parametrize residual gauge symmetries through {\em field dependent
  redefinitions}. In a first step, one trades the real function
$\d_uf(u, \zeta, \bar{\zeta})$ and the imaginary
$\omega'^{34}_0(u, \zeta, \bar{\zeta})$ for a complex
$\Omega(u, \zeta, \bar{\zeta})$ according to
\begin{equation}
\begin{split}
\partial_u f &= \half[\bar\d\bar Y-\bar Y\bar \d\ln (P\bar P) + \d Y
-Y\d \ln (P\bar P)] + f
(\gamma^0+\xbar\gamma^0)+\half(\Omega+\bOmega), \\
\omega'^{34}_0 &= \half[\bar\d \bar Y-\bar Y\bar \d\ln P +\bar Y \bar \d
\ln \bar P- \d Y + Y\d \ln \bar P -Y \d \ln P]\\ &\hspace{8cm}+
f(\xbar\gamma^0 - \gamma^0)+
\half(\Omega-\bOmega).\\
\end{split}
\label{parameter redef 4d}
\end{equation}
It then follows that the first of \eqref{parameter redef 4d} can be
solved for $f$ in terms of an
integration function $T_R(\zeta,\bar\zeta)$, (called $\tilde T$ in
\cite{Barnich:2010eb,Barnich:2011mi,Barnich:2013oba})
\begin{equation}
f(u, \zeta, \bar{\zeta}) = \frac{1}{\sqrt{P\bar{P}}} [T_R(\zeta,
\bar{\zeta}) + \frac{\tilde{u}}{2} (\partial Y + \bar{\partial}
\bar{Y}) - Y \partial \tilde{u} - \bar{Y} \bar{\partial} \tilde{u} +
\frac{1}{2} (\tilde{\Omega} + \bar{\tilde{\Omega}})]
\label{explicit f},
\end{equation}
where
\begin{equation}
\tilde{u} = \int^u_{u_0} du' \sqrt{P\bar{P}}, \quad \tilde{\Omega} =
\int^u_{u_0} du' \sqrt{P\bar{P}}\,\Omega.
\end{equation}
This redefinition of parameters is such that
\begin{equation}
\boxed{-\delta_{Y,T_R,\Omega} P = \Omega P} ,
\label{transfo finale}
\end{equation}
together with the complex conjugate relation
$-\delta_{Y,T_R,\Omega} \bar{P} = \bar{\Omega}\bar{P}$.

Denoting by $\phi^\alpha$ the fields
$({e_a}^\mu,\Gamma_{abc})$ (together with the auxiliary fields ${\bf
R}_{abcd},\lambda^{abcd}$ if useful), it follows from \eqref{eq:15}
that
\begin{equation}
  \begin{split}
& [\delta_{\xi'_1,\omega'_1},\delta_{\xi'_2,\omega'_2}]\phi^\alpha=
-\delta_{\hat\xi',\hat\omega'}\phi^\alpha,\\
& \hat\xi'^\mu=[\xi'_1,\xi'_2]^\mu, \quad (\hat{\omega'})\indices{_a^b}
  ={\xi'_1}^\rho\p_\rho{\omega'_{2a}}^b+{\omega'_{1a}}^c{\omega'_{2c}}^b
  -(1\leftrightarrow
  2),
\end{split}
\end{equation}
when the gauge parameters $\xi',\omega'$ are field-independent. In
case gauge parameters do depend on the fields, one finds instead
\begin{equation}
  \begin{split}
& [\delta_{\xi'_1,\omega'_1},\delta_{\xi'_2,\omega'_2}]\phi^\alpha=
-\delta_{\hat\xi'_M,\hat\omega'_M}\phi^\alpha,\\
& \hat\xi'^\mu_M=[\xi'_1,\xi'_2]^\mu
-\delta_{\xi'_1,\omega'_1}{\xi'}^\mu_2+\delta_{\xi'_2,\omega'_2}{\xi'}^\mu_1,
  \\
 & (\hat{\omega'}_M)\indices{_a^b}
 ={\xi'_1}^\rho\p_\rho{\omega'_{2a}}^b+{\omega'_{1a}}^c{\omega'_{2c}}^b
 -\delta_{\xi'_1,\omega'_1}{\omega'_{2a}}^b
  -(1\leftrightarrow
  2).
\end{split}
\end{equation}
We now have the following result:

The gauge parameters $(\xi'[Y,T_R,\Omega],\omega'[Y,T_R,\Omega])$
equipped with the modified commutator for field dependent gauge
transformations realize the direct sum of the abelian ideal of complex
Weyl rescalings with the (extended) $\mathfrak{bms}_4$ algebra
everywhere in the bulk spacetime,
\begin{equation}
  \label{eq:105}
  \begin{split}
  &\hat \xi'_M=\xi'[\hat Y,\hat T_R,\hat \Omega],\quad \hat
  \omega'_M=\omega'[\hat Y,\hat T_R,\hat \Omega],\\
  & \hat Y^A=Y_1^B\partial_B Y^A_2-Y_2^B\partial_B Y^A_1,\\
  &\hat T_R=Y_1^A\d_A T_{R2}+\half T_{R1}\d_A Y^A_2 -(1\leftrightarrow
  2),\\
  & \hat\Omega=0.
\end{split}
\end{equation}
The proof follows by adapting the ones provided in
\cite{Barnich:2010eb,Barnich:2013oba,Barnich:2013sxa} to the current
set-up.

\subsection{Action of symmetries on solutions}
\label{sec:acti-symm-solut-1}

A further field-dependent redefinition consists in defining
\begin{equation}
  \label{eq:95}
  Y =\bar P \xbar \cY,\quad \bar Y =P \cY,
\end{equation}
where the spin weights of $\xbar \cY$ and $\cY$ are given
  in Table \ref{t1}. These quantities are
  more convenient when using the operators $\eth$ and $\bar\eth$. The action of asymptotic symmetries
  on solutions is given in the original parametrization in Appendix \ref{sec:oracti}. In terms of
  the redefined parameters, the transformations \eqref{transfo solution space 1} then become
\begin{equation}
\begin{split}
&-\delta_s \sigma^0=[\cY \eth +\xbar \cY \xbar\eth +
\frac32 \eth\cY - \frac12 \xbar\eth\xbar\cY + \frac{3}{2}\Omega -
\frac{1}{2}\bar{\Omega}]\sigma^0 + f\bar\lambda^0 - \eth^2 f,\\
&-\delta_s \Psi^0_0=[\cY \eth +\xbar \cY \xbar\eth +
\frac52 \eth\cY + \frac12\xbar\eth\xbar\cY +\frac{5}{2}\Omega +
\frac{1}{2}\bar{\Omega}]\Psi^0_0 + f\eth\Psi_1^0 + 3f\sigma^0\Psi_2^0
+ 4 \Psi^0_1 \eth f,\\
&-\delta_s \Psi^0_1=[\cY \eth +\xbar \cY \xbar\eth + 2
\eth\cY + \xbar\eth\xbar\cY + 2\Omega + \bar{\Omega}]\Psi^0_1 +
f\eth\Psi_2^0 + 2f\sigma^0\Psi_3^0 + 3 \Psi^0_2 \eth f,\\
&-\delta_s \left(\frac{\Psi^0_2 +
    \bar{\Psi}^0_2}{2}\right)=[\cY \eth +\xbar \cY \xbar\eth + \frac32
\eth\cY + \frac32\xbar\eth\xbar\cY  + \frac{3}{2}\Omega +
\frac{3}{2}\bar{\Omega}]\left(\frac{\Psi^0_2 +
    \bar{\Psi}^0_2}{2}\right)\\
&~~~~~~~~~~~~~~~~~~~~~~~~~~~~~~ + \frac{1}{2}( f\eth\Psi_3^0
+f\sigma^0\Psi_4^0+ 2 \Psi^0_3 \eth f + \text({c.c.})),
\end{split}
\label{transfo final 2}
\end{equation}
while \eqref{transfo solution space 2}-\eqref{transfo solution space
  4} read as
\begin{multline}
-\delta_s \Psi^1_0 =\big[\cY \eth +\xbar \cY \xbar\eth
+ 3 \eth\cY + \xbar\eth\xbar\cY
+3\Omega + \bar{\Omega}\big]\Psi^1_0\\ - \overline{\eth}\big[5  \eth f
\Psi^0_0 + f \eth \Psi^0_0 + 4 f \Psi^0_1 \sigma^0 \big],
\end{multline}
\begin{multline}
  -\delta_s \Psi^2_0 = \big[\cY \eth +\xbar \cY \xbar\eth
+\frac{7}{2} \eth\cY +\frac{3}{2} \xbar\eth\xbar\cY +\frac{7}{2}\Omega
+ \frac{3}{2}\bar{\Omega}\big] \Psi^2_0 \\ + \big[-3 \bar{\Delta} f -
\xbar \eth f \eth - 3 \eth f \overline{\eth} -
\frac{1}{2} f \eth \overline{\eth} - \frac{5}{4} f R \big] \Psi^1_0
\\ +\big[-5 f \Psi^0_2 - \frac{5}{2} f \bar{\Psi}^0_2 + \frac{5}{2} f
\sigma^0 \overline{\eth}^2 + 5f \overline{\eth} \sigma^0
\overline{\eth} + 3 f \eth \bar{\sigma}^0 \eth + \frac{1}{2} f
\bar{\sigma}^0 \eth^2 + \frac{5}{2} f \eth^2 \bar{\sigma}^0 +
\frac{5}{2}f \sigma^0
\lambda^0 \\ + 5 \overline{\eth} \sigma^0 \overline{\eth} f + 15 \eth
\bar{\sigma}^0 \eth f + 5 \sigma^0 \overline{\eth} f \overline{\eth} +
3 \bar{\sigma}^0 \eth f \eth \big] \Psi^0_0
\\ +\big[ 5 f \Psi_1^0 + 12 \sigma^0 \bar{\sigma}^0 \eth f  + 12 f \sigma^0
\eth \bar{\sigma}^0 + 2f \eth \sigma^0 \bar{\sigma}^0 + \frac{9}{2}f
\sigma^0 \bar{\sigma}^0 \eth  \big] \Psi^0_1 \\
+ \frac{15}{2} f(\sigma^0)^2 \bar{\sigma}^0 \Psi^0_2
\end{multline}
and, by induction,
\begin{multline}
-\delta_s \Psi^n_0 =\big[ \cY \eth +\xbar \cY \xbar\eth +
\frac{5+n}{2} \eth\cY + \frac{1+n}{2}
\xbar\eth\xbar\cY + \frac{5+n}{2} \Omega +
\frac{1+n}{2} \bar{\Omega}\big]\Psi^n_0 \\
+ (\text{inhomogeneous terms}).
\label{transfo final 3}
\end{multline}
Finally, the variations \eqref{transfo
  solution space} are given by
\begin{equation}
\begin{split}
&-\delta_s \lambda^0=[\cY \eth +\xbar \cY \xbar\eth +2
\xbar\eth\xbar\cY + 2 \bar{\Omega}]\lambda^0 - f \Psi_4^0 -
\frac12\xbar\eth^2(\eth\cY+\xbar\eth\xbar\cY),\\
&-\delta_s \Psi^0_2=[\cY \eth +\xbar \cY \xbar\eth +
\frac32 \eth\cY + \frac32\xbar\eth\xbar\cY  + \frac{3}{2}\Omega +
\frac{3}{2}\bar{\Omega}]\Psi^0_2 \\ &\hspace{8cm} + f\eth\Psi_3^0 +f\sigma^0\Psi_4^0+ 2
\Psi^0_3 \eth f,\\
&-\delta_s \Psi^0_3=[\cY \eth +\xbar \cY \xbar\eth +
\eth\cY + 2\xbar\eth\xbar\cY+ \Omega +2\bar{\Omega}]\Psi^0_3 +
f\eth\Psi_4^0 + \Psi^0_4 \eth f,\\
&-\delta_s \Psi^0_4=[\cY \eth +\xbar \cY \xbar\eth +
\frac12 \eth\cY + \frac52\xbar\eth\xbar\cY +\frac{1}{2}\Omega +
\frac{5}{2}\bar{\Omega}]\Psi^0_4 \\ &\hspace{8cm} + f \p_u \Psi_4^0 +
2(2\gamma^0+\xbar\gamma^0)\Psi_4^0.
\end{split}
\label{transfo finale v2}
\end{equation}

\subsection{Reduction of solution space}
\label{sec:reduct-solut-space}

Besides conditions \eqref{gauge conditions} and \eqref{fall-off
  conditions}, additional constraints may be imposed on solution
space. A standard choice is to fix the conformal factor $P$ to be
equal to $P_S$ given in \eqref{unit}. We will also fix $P$ here,
without committing to a specific value. In other words, we consider
$P$ to be part of the background structure. As a consequence,
infinitesimal complex Weyl rescalings (whose finite counterparts have
been discussed in \cite{Barnich:2016lyg}) are frozen and $\Omega=0$ in
the formulas above, while in the formulas below, $s$ stands for
$(\cY,\xbar\cY,T_R,0)$. The main reason why we do not perform the
analysis below while keeping $P(u,\zeta,\bar\zeta)$ arbitrary is
computational simplicity. We will return elsewhere to a detailed
discussion of the current algebra and its interpretation when complex
Weyl rescalings are allowed.

\subsection{Breaking and co-dimension 2 form}
\label{sec:break-co-dimens}

Under this additional constraint on solution space, the breaking
(and thus also the invariant presymplectic $(3,2)$ form) can be
computed using equation \eqref{eq:25a},
\begin{equation}
\star b_{s} = - b^r_{s(0)} du d\zeta d\bar{\zeta} +
\mathcal{O}(r^{-1})\label{eq:98},
\end{equation}
where
\begin{equation}
b^r_{s (0)}=\frac{1}{8\pi G
  P\bP}\left(\delta\sigma^0\delta_s\lambda^0 +
  \delta\xbar\sigma^0\delta_s\xbar\lambda^0 -
  \delta\lambda^0\delta_s\sigma^0 -
  \delta\xbar\lambda^0\delta_s\xbar\sigma^0\right).
\label{breaking 4d}
\end{equation}
Since the breaking contains the
  information about the non-conservation of the currents, it should
  not come as a surprise that it depends on the news functions encoded
  in $\lambda^0$ and $\bar{\lambda}^0$. Furthermore, the co-dimension
$2$ form \eqref{eq:130} takes the form
\begin{equation}
\star k_{s} = k^{ur}_{s(0)}  d\zeta d\bar{\zeta} - k^{\zeta
  r}_{s(0)} du  d\bar{\zeta} + k^{\bar{\zeta}r}_{s(0)} du  d\zeta +
\mathcal{O}(r^{-1})
\end{equation} where
\begin{multline}
k^{ur}_{s(0)}=-\frac{1}{P\bP 8\pi G}\Big(\delta\big[f(\Psi^0_2 + \sigma^0
\lambda^0) +\cY(\sigma^0\eth\xbar\sigma^0  + \half
\eth(\sigma^0\xbar\sigma^0)+ \Psi^0_1) \\- \half \eth(\cY
\sigma^0\xbar\sigma^0) - r \eth(\xbar\cY \xbar\sigma^0)\big]  - f
\lambda^0 \delta \sigma^0 + c.c.\Big),\label{current-full}
\end{multline}
\begin{multline}
k^{\zeta r}_{s(0)}=-\frac{1}{P 8\pi
  G}\Big(\delta\big[\xbar\cY(\xbar\lambda^0\xbar\sigma^0 - \xbar
\Psi^0_2) - f\xbar\Psi^0_3 + \half \xbar\eth\sigma^0(\eth \cY -
\xbar\eth\xbar\cY)+\half \sigma^0\xbar\eth(\xbar\eth\xbar\cY -
\eth\cY)\\ - \xbar\lambda^0\xbar\eth f + r \cY (\xbar\lambda^0 +
\sigma^0(\gamma^0+\xbar\sigma^0))\big]-\xbar\cY(\xbar\lambda^0\delta\xbar\sigma^0
+ \lambda^0\delta\sigma^0) \Big),
\end{multline}
and $k^{\bar\zeta r}_{s(0)}$ given by the complex conjugate.
By construction
\begin{equation}
\p_u k^{ur}_{s(0)} + \p_\zeta k^{\zeta r}_{s(0)} +
\p_{\bar\zeta} k^{\bar\zeta r}_{s(0)}=-b^r_{s(0)}\label{eq:97},
\end{equation}
which may also be checked by direct computation. Note that
$k^{ur}_{s(0)},k^{\zeta r}_{s(0)},k^{\bar\zeta r}_{s(0)}$ contain,
besides a finite contribution, also linearly divergent terms when
$r\to\infty$.  Following \cite{Barnich:2013axa}, the latter can be
removed through an exact $2$-form $\p_\rho
\eta_s^{\mu\nu\rho}$. Defining
\begin{equation}
\bP\eta_s^{[ur\bar{\zeta}]}=\cN_s^u=-r\xbar\cY \xbar\sigma^0 - \half \cY
\sigma^0\xbar\sigma^0,\quad
\eta_s^{[\zeta r\bar{\zeta}]}=\cN^\zeta_s=0,
\end{equation}
and splitting into
an integrable part
\begin{equation}
\boxed{\cJ^u_s=-\frac{1}{8\pi
  G}\big[f(\Psi^0_2+ \sigma^0\lambda^0) + \cY[\sigma^0\eth\xbar\sigma^0 +
\Psi^0_1 + \half \eth(\sigma^0\xbar\sigma^0)]+ \text{c.c.}\big]}, \label{current 1}
\end{equation}
\begin{multline}
  \cJ^\zeta_s=\frac{1}{8\pi G}\bigg[\xbar\cY \xbar
  \Psi^0_2 + f\xbar\Psi^0_3 +\half \xbar\cY(\lambda^0\sigma^0 -
  \xbar\lambda^0\xbar\sigma^0) + \half
  \xbar\eth\sigma^0(\xbar\eth\xbar\cY - \eth \cY)\\ - \half
  \sigma^0\xbar\eth(\xbar\eth\xbar\cY - \eth\cY) +
  \xbar\lambda^0\xbar\eth f\bigg], \label{current 2}
\end{multline}
and a non-integrable one
\begin{equation}
\Theta^u_s(\delta\chi)=\frac{1}{8\pi G}(f\lambda^0\delta\sigma^0 + \text{c.c.}),\quad
\Theta^\zeta_s(\delta\chi)=\frac{1}{8\pi G}\xbar\cY(\lambda^0\delta\sigma^0 +
\xbar\lambda^0\delta\xbar\sigma^0), \label{current 3}
\end{equation}
one finally arrives at
\begin{equation}
  \begin{split}
\delta \cJ^u_s&=P\bP[k^{ur}_{s(0)} - \bp \eta_s^{[ur\bar{\zeta}]} - \p
\eta_s^{[ur\zeta]}] - \Theta^u_s,\\
\delta \cJ_s^\zeta&=P [k^{\zeta r}_{s(0)} + \p_u \eta_s^{[ur\zeta]}
+\bp\eta_s^{[\bar{\zeta} r \zeta]}]- \Theta_s^\zeta,
\end{split}
\end{equation}
where $\cJ_s^{\bar{\zeta}},\Theta_s^{\bar{\zeta}}$ are the complex
conjugates of $\cJ_s^{\zeta},\Theta_s^{\zeta}$.
  Expressions \eqref{current 1}, \eqref{current 2} and \eqref{current
    3} are the final results for the BMS currents. Notice that the
  split between integrable and non-integrable part is ambiguous and,
  as shown in the next subsection, it is crucial to keep track of both
  parts. The results of \cite{Barnich:2013axa} are recovered when
taking $P$ to be $u$-independent, which implies
$\gamma^0=\xbar\gamma^0=0$ and $\lambda^0=\dot{\xbar\sigma^0}$. Note
that the associated forms are given by
\begin{equation}
  \label{forms}\begin{split}
J_s &= (P\bar P)^{-1}\cJ^u_s d\zeta d\bar{\zeta} - P^{-1}\cJ^\zeta_s du d\bar{\zeta} +
\bar P^{-1} \cJ^{\bar{\zeta}}_s du d\zeta, \\
\theta_s &= (P\bar P)^{-1}\Theta^u_s d\zeta d\bar{\zeta} - P^{-1}\Theta_s^{\zeta} du
d\bar{\zeta} + \bar P^{-1} \Theta_s^{\bar{\zeta}} du d\zeta.
\end{split}
\end{equation}

\subsection{Current algebra}
\label{sec:algebra}

Even if the co-dimension $2$ form derived in the
  previous subsection leads to non-integrable expressions, one can
  still define a consistent current algebra whose general structure
  does not depend on the particular split between integrable and
  non-integrable pieces \cite{Barnich:2013axa,Barnich:2011mi}. As
  briefly recalled in section \ref{sec:discussion}, this algebra
  contains important information on physical properties of the
  system. Using the relations of Appendix \ref{Useful relations}, the
first independent component of the current algebra can be written as
\begin{equation}
  \boxed{-\delta_{s_2} \cJ^u_{s_1} +
    \Theta_{s_2}^u(-\delta_{s_1}\chi)\approx
    \cJ^u_{[s_1,s_2]}+\cK^u_{s_1,s_2}+\eth\cL_{s_1,s_2}+\xbar\eth\xbar{\cL_{s_1,s_2}}},
  \label{current algebra 1}
\end{equation}
where
\begin{equation}
\cK^u_{s_1,s_2}=\frac{1}{8\pi G}\Big[\Big(\frac12\xbar\sigma^0\left[f_1 \eth^2(\eth\cY_2 +
  \xbar\eth\xbar\cY_2)\right] - f_1\eth f_2 \xbar\eth\mu^0
-(1\leftrightarrow 2)\Big)+ \text{c.c.}\Big],
\end{equation}
and
\begin{multline}
\cL_{s_1,s_2}=\cY_2 \cJ^u_{s_1} - f_2 \cJ^{\bar{\zeta}}_{s_1}\\
-\frac{1}{8\pi G}\Big[\big(
\frac12 (\eth\cY_1 + \xbar\eth\xbar\cY_1) \eth f_2 -\frac12
\cY_1\eth^2f_2 - \xbar\cY_1\eth\xbar\eth f_2\big)\xbar\sigma^0 \\ -
\frac12 \cY_1 \xbar\eth^2f_2 \sigma^0 - \cY_1\eth f_2\eth\xbar\sigma^0
+ \xbar\cY_1 \xbar\eth f_2\eth \xbar\sigma^0 - f_1\eth f_2
\lambda^0\Big].
\end{multline}
The second independent component of the current algebra is
\begin{equation}
\boxed{-\delta_{s_2} \cJ^{\bar{\zeta}}_{s_1} +
\Theta_{s_2}^{\bar{\zeta}}(-\delta_{s_1}\chi)
\approx \cJ^{\bar{\zeta}}_{[s_1,s_2]}+\cK^{\bar{\zeta}}_{s_1,s_2}
-\p_u\cL_{s_1,s_2}-2\gamma^0\cL_{s_1,s_2}+\xbar\eth\xbar{
\cM_{s_1,s_2}}},\label{eq:101}
\end{equation}
where
\begin{multline}
\cK^{\bar{\zeta}}_{s_1,s_2}=-\frac{1}{8\pi G}\Big[f_2\eth f_1 \xbar\eth\nu^0 + \frac12 \eth
f_1\xbar\eth^3\xbar\cY_2 + \cY_1\xbar\eth f_2 \eth \mu^0 + f_1
\cY_2(\sigma^0\xbar\eth\nu^0 + \xbar\sigma^0\eth\xbar\nu^0) \\+
\frac12\cY_2\xbar\eth^2(\eth\cY_1 + \xbar\eth\xbar\cY_1)\sigma^0 +
\frac12\cY_2\eth^2(\eth\cY_1 + \xbar\eth\xbar\cY_1)\xbar\sigma^0 -
(1\leftrightarrow2)\Big],
\end{multline}
and
\begin{equation}
\xbar{\cM_{s_1,s_2}}=\xbar\cY_2 \cJ^{\bar{\zeta}}_{s_1}
-\frac{1}{8\pi G}\Big[\frac12\xbar\eth(\xbar\eth\xbar\cY_1 - \eth\cY_1 )\eth f_2 + \frac12
\eth \cY_1 \eth\xbar\eth f_2\Big] - \text{c.c.}.\label{eq:102}
\end{equation}

\subsection{Cocycle condition}
\label{sec:cocycle-condition}

The components of $\cK_{s_1,s_2}$
satisfy the 2-cocycle conditions
\begin{equation}
\cK^u_{[s_1,s_2],s_3}-\delta_{s_3} \cK^u_{s_1,s_2} +
\text{cyclic}(1,2,3)=\eth\cN_{s_1,s_2,s_3}+\xbar\eth\xbar\cN_{s_1,s_2,s_3},
\end{equation}
where
\begin{equation}
\cN_{s_1,s_2,s_3}=-f_3 \cK^{\bar{\zeta}}_{s_1,s_2}+ \text{cyclic}(1,2,3),
\end{equation}
and
\begin{equation}
\cK^{\bar{\zeta}}_{[s_1,s_2],s_3}-\delta_{s_3}
\cK^{\bar{\zeta}}_{s_1,s_2} + \text{cyclic}(1,2,3)=-\p_u\cN_{s_1,s_2,s_3} - 2
\gamma^0 \cN_{s_1,s_2,s_3} + \xbar\eth\xbar{\cO_{s_1,s_2,s_3}},
\end{equation}
where
\begin{multline}
\xbar{\cO_{s_1,s_2,s_3}}=-\frac{1}{8\pi G}\xbar\cY_3\Big[(f_1\cY_2-f_2\cY_1)\sigma^0\xbar\eth\nu^0 +
\frac12\sigma^0(\cY_2\xbar\eth^3\xbar\cY_1 -
\cY_1\xbar\eth^3\xbar\cY_2) \\+ \frac12(\eth f_1\xbar\eth^3\xbar\cY_2
- \eth f_2\xbar\eth^3\xbar\cY_1) + (f_2\eth f_1 - f_1\eth
f_2)\xbar\eth\nu^0\Big]-\text{c.c.} + \text{cyclic}(1,2,3).
\end{multline}
A situation where this 2-cocycle is relevant is discussed in
\cite{Barnich:2017ubf}.

\section{Discussion}
\label{sec:discussion}

Let us briefly recall the discussion in
\cite{Barnich:2013axa,Barnich:2011mi} on the physical interpretation
of BMS charge and current algebras.

When one restricts to globally well-defined quantities on the
sphere, with $P=P_S=\frac{1}{\sqrt{2}}(1+\zeta\bar\zeta)$, there are
no superrotations and $\cK^u_{s_1,s_2}=0=\cK^{\bar\zeta}_{s_1,s_2}$. In
this case, BMS charges
are defined by integrating the forms \eqref{forms} at fixed retarded
time over the celestial sphere,
\begin{equation}
Q_s=\int_{u={\rm cte}} J_s=\int_{u={\rm cte}} (P_S\bar
P_S)^{-1}\cJ^u_s d\zeta d\bar\zeta.\label{eq:86}
\end{equation}
If one also defines
\begin{equation}
  \label{eq:120}
  \Theta_{s}=\int_{u={\rm cte}} \theta_s=\int_{u={\rm cte}} (P_S\bar
P_S)^{-1}\Theta^u_s d\zeta d\bar\zeta,
\end{equation}
and the bracket
\begin{equation}
\{Q_{s_1},Q_{s_2}\}^*=-\delta_{s_2}
Q_{s_1}+\Theta_{s_2}[-\delta_{s_1}\chi], \label{eq:113}
\end{equation}
the integrated version
of equation \eqref{current algebra 1},
becomes
\begin{equation}
  \label{eq:116}
  \{Q_{s_1},Q_{s_2}\}^*=Q_{[s_1,s_2]},
\end{equation}
This charge algebra contains for instance the information on
non-conservation of BMS charges. Indeed, let us take for
 $s_2=\d_u$ by which we mean that $T_R=\sqrt{P_S\bar P_S}, Y=0=\bar Y$, so that
  $f=1,\cY=0=\bar\cY$. In this case, equation \eqref{eq:116} together with the
  definition of the left hand side in \eqref{eq:113}
  becomes
  \begin{equation}
    \label{eq:118}
    -\delta_{\d_u} Q_s+\Theta_{\d_u}[-\delta_s\chi]=Q_{[s,\d_u]}.
  \end{equation}
  When using that
  \begin{equation}
    \label{eq:117}
    \frac{d}{du} Q_s=-\delta_{\d_u} Q_s+\dover{}{u}Q_s,
  \end{equation}
  and $\dover{}{u}Q_s =Q_{{\d s}/{\d u}}=-Q_{[s,\d_u]}$, it follows
  that
\begin{equation}
  \frac{d}{du} Q_s= \Theta_{\d_u}[\delta_{s_1}\chi].
\end{equation}
If one now chooses $s=\d_u$, one recovers the Bondi mass loss
formula.

More generally, equation \eqref{current algebra 1} is the local
version of \eqref{eq:116} where superrotations and arbitary fixed
$P(u,\zeta,\bar\zeta)$ are allowed. When choosing $s_2=\d_u$ in that
equation, it encodes the non-conservation of BMS currents
(cf.~equation (4.22) of \cite{Barnich:2013axa}). For particular
choices of $s_1$, it controls the time evolution of the Bondi mass and
angular momentum aspects.

Even though we concentrated here on the case of standard Einstein
gravity, all the kinematics is in place to generalize the
constructions to gravitational theories with higher derivatives and/or
dynamical torsion.

For most part of the paper, the standard discussion has been extended
so as to include an arbitrary u-dependent conformal factor $P$. This
has been done so as to manifestly include the Robinson-Trautman
solution \cite{Robinson:1960zzb,Robinson:1962zz} in solution
space. The application of the current set-up to these solutions
requires the inclusion of a dynamical conformal factor in the
derivation of the current algebra. We plan to address this question
elsewhere.

\section*{Acknowledgements}
\label{sec:acknowledgements}

\addcontentsline{toc}{section}{Acknowledgments}

The work of G.B.~and of R.R.~is supported by the F.R.S.-FNRS Belgium,
convention FRFC PDR T.1025.14 and convention IISN 4.4503.15. The work
of P.~Mao is supported in part by the China Postdoctoral Science Foundation
under Grant No. 2017M620908 and the National Natural Science Foundation
of China under Grant Nos. 11905156, 11935009, and 11575202.

\appendix

\section{(Non)-conservation of codimension 2 forms in first order
  gauge theories}
\label{sec:non-cons-codim}

In order to prove equation \eqref{eq:85}, we need in a first step to
work out all consequences of the Noether identities \eqref{eq:149} for
first order gauge theories.

In the context of variational calculus, off-shell identities between
the fields and their derivatives have to hold for all possible values
of these variables. In other words, the fields and their derivatives
are considered as independent coordinates on a suitable space, the
so-called jet-space. It thus follows that the Noether identities give
rise to separate identities when considering terms involving
$\d_\mu\d_\nu\phi^j$, $\d_\mu\phi^k\d_\nu\phi^j$, $\d_\mu\phi^j$ or no
derivatives. The Noether identities are thus equivalent to
\begin{equation}
  \label{eq:150}
  \begin{split}
    & R^{i(\mu}_\alpha \sigma_{ij}^{\nu)}=0,\\
    & \d_k(R^{i
      \mu}_\alpha\sigma^\nu_{ij})+\d_j(R^{i\nu}_\alpha\sigma^\mu_{ik})
    +R^{i\mu}_{k\alpha}\sigma^\nu_{ij}+R^{i\nu}_{j\alpha}\sigma^\mu_{ik}=0,\\
    &R^{i0}_\alpha\sigma^\mu_{ij}+\d_j[R^{i\mu}_\alpha(\d_ih+\dover{}{x^\nu}
    a^\nu_i)]-R^{k\mu}_{j\alpha}(\d_k h+\dover{}{x^\nu}
    a^\nu_k)-\dover{}{x^\nu} (R^{i\nu}_\alpha\sigma_{ij}^\mu)=0,\\
    & R^{i0}_\alpha(\d_i h+\dover{}{x^\nu}
    a^\nu_i)-\dover{}{x^\mu}[R^{i\mu}_\alpha(\d_i h+\dover{}{x^\nu}
    a^\nu_i)]=0.
  \end{split}
\end{equation}
As discussed above, the linearized equations of motion
derive from the action
  \begin{equation}
    \label{eq:152}
    L^{(2)}[\varphi,\phi]=\d_i a_j^\mu\varphi^i\d_\mu\varphi^j +\half
    \d_i\d_j a^\mu_k\varphi^i \varphi^j\d_\mu\phi^k-\half \d_i\d_j
    h\varphi^i\varphi^j,
  \end{equation}
  so that the left hand sides of the linearized equations of motion are
  given by
\begin{equation}
  \label{eq:151}
  \vddl{L^{(2)}[\varphi,\phi]}{\varphi^i}=  [\sigma^\mu_{ij}\d_\mu
  +\d_j\sigma^\mu_{ik}\d_\mu\phi^k-\d_j(\d_i h+ \dover{}{x^\nu}
    a^\nu_i)]\varphi^j.
\end{equation}
Let us then explicitly work out $\d_\nu k^{[\mu\nu]}_f$ with
$k^{[\mu\nu]}_f$ given in \eqref{eq:81} by controlling the derivatives
of the fields and the gauge parameters that appear. By using the first
of \eqref{eq:150}, it follows that
\begin{equation}
 \d_\nu  k^{[\mu\nu]}_f=R^{i\mu}_\alpha
   \sigma^{\nu}_{ij}\d_\nu\varphi^j f^\alpha-R^{i\nu}_\alpha
   \sigma^\mu_{ij}\varphi^j\d_\nu
   f^\alpha+\d_\nu(R^{i\mu}_\alpha\sigma^\nu_{ij})\varphi^jf^\alpha.
 \end{equation}
In the first term, we eliminate $\sigma^{\nu}_{ij}\d_\nu\varphi^j$ in
terms of undifferentiated $\varphi^j$ by using the linearized
equations of motion. In the second term , we write
$-R^{i\nu}_\alpha\partial_\nu
f^\alpha$ as $-R^i_\alpha[f^\alpha]+R^i_\alpha f^\alpha$. In the last
term, we have
$\d_\nu(R^{i\mu}_\alpha\sigma^\nu_{ij})=\dover{}{x^\nu}
(R^{i\mu}_\alpha\sigma^\nu_{ij})+\d_k(R^{i\mu}_\alpha\sigma^\nu_{ij})\d_\nu\phi^k$,
and we then use the second of \eqref{eq:150} to re-write the last term
of this expression.
We then have
\begin{multline}
  \label{eq:114}
  \d_\nu  k^{[\mu\nu]}_f+
  W^\mu_{{\delta\cL}/{\delta\phi}}[\varphi,R_\alpha[f^\alpha]] =
  \big(R^{i\mu}_\alpha [-\d_j\sigma^\nu_{ik}\d_\nu\phi^k+\d_j(\d_i h+ \dover{}{x^\nu}
    a^\nu_i)] +R^i_\alpha \sigma^\mu_{ij}\\ +\dover{}{x^\nu}
(R^{i\mu}_\alpha\sigma^\nu_{ij})-[\d_j(R^{i\nu}_\alpha\sigma^\mu_{ik})
    +R^{i\mu}_{k\alpha}\sigma^\nu_{ij}+R^{i\nu}_{j\alpha}\sigma^\mu_{ik}]\d_\nu\phi^k\big)\varphi^j
    f^\alpha.
  \end{multline}
  In the 2nd term on the last line, we may write
  $ -\d_j(R^{i\nu}_\alpha\sigma^\mu_{ik})\d_\nu\phi^k=\d_j
  R^{i\mu}_\alpha\sigma^\nu_{ik}\d_\nu\phi^k+R^{i\mu}_\alpha\d_j
  \sigma^\nu_{ik}\d_\nu\phi^k$ by using again the first of
  \eqref{eq:150}. The last of these terms then vanishes with the first
  one on the right hand side of \eqref{eq:114}, whereas for the first
  of these
  terms, we may use the full equations of motion \eqref{eq:148} to eliminate
  $\sigma^\nu_{ik}\d_\nu\phi^k$. When using in addition \eqref{eq:94}
  to simplify the last term of the first line and the last term of
  the last line of
  \eqref{eq:114}, the right hand side of \eqref{eq:114} reduces to
 \begin{equation*}
   [R^{i\mu}_\alpha\d_j(\d_i h+ \dover{}{x^\nu}
    a^\nu_i)+R^{i0}_\alpha \sigma^\mu_{ij}+\dover{}{x^\nu}
(R^{i\mu}_\alpha\sigma^\nu_{ij})+\d_j R^{i\mu}_\alpha(\d_i
h+\dover{a_i}{x^\nu})-R^{i\mu}_{k\alpha}\sigma^\nu_{ij}\d_\nu\phi^k]\varphi^jf^\alpha
 \end{equation*}
When using the first of \eqref{eq:150} for the term $\dover{}{x^\nu}
(R^{i\mu}_\alpha\sigma^\nu_{ij})$ and the full equations of motion to
eliminate $\sigma^\nu_{ij}\d_\nu\phi^k$ in the last term, this
expression reduces
to the left hand side of the third of \eqref{eq:150} and thus vanishes
identically.

\section{Frames and forms}
\label{sec:conventions-forms}

\subsection{Frames and directional derivatives}
\label{sec:fram-direct-deriv}

Consider an $n$-dimensional spacetime with a moving frame
${e_a}^\mu{e^a}_\nu=\delta^\mu_\nu$,
${e_a}^\mu{e^b}_\mu=\delta_a^b$. Let
\begin{equation}
e_a={e_a}^\mu\d_\mu,\quad {}^*e^{a}={e^a}_\mu dx^\mu, \label{eq:2}
\end{equation}
and $\d_a f={e_a}^\mu\d_\mu f$.  The structure functions are defined
by
\begin{equation}
[e_a,e_b]={D^c}_{ab}e_c \iff d{}^*e^{a}=-\half {D^a}_{bc}{}^*e^{b}{}^*e^{c}.\label{eq:13}
\end{equation}
If one defines
\begin{equation}
  \label{eq:49}
   {d^a}_{bc}={e^a}_\mu \d_b {e_c}^\mu,
\end{equation}
then
\begin{equation}
{d^\mu}_{\nu\lambda}=-{e_d}^\mu\d_\nu {e^d}_\lambda,\quad
{D^a}_{bc}=2{d^a}_{[bc]}, \label{eq:100}
\end{equation}
where it is understood that tangent space indices $a,b,\dots$ and
world-indices $\mu,\nu,\dots$ are transformed into each other by using
the vielbeins and their inverse.
For later use, note that if ${\mathbf e}={\rm det}\,{e^a}_\mu$, then
\begin{equation}
  \label{eq:82}
  \d_\mu(\mathbf{e}\,{e^\mu}_a)=\mathbf{e}\, {D^b}_{ba}.
\end{equation}

\subsection{Horizontal complex}
\label{sec:horizontal-complex}

The differential forms
$\omega=\sum^n_{k=0}\frac{1}{k!}\omega_{a_1\dots a_k} {}^*e^{a_1}\dots
{}^*e^{a_k}$ that are useful for our purpose are ``local forms''. They
can be considered as polynomials in the independent, anticommuting
variables ${}^*e^{a}$ (i.e., the wedge product is omitted), with
coefficients that depend on $x^\mu$, and the fields $\phi^i$ (that
include ${e_a}^\mu$ together with the other relevant fields), and a
finite number of their derivatives, considered as independent
variables. In this context,
$\d_\mu=\ddl{}{x^\mu}+\phi^i_{,\mu}\ddl{}{\phi^i}+\dots$ is the
horizontal derivative of the variational bicomplex (see
e.g.~\cite{Anderson1991,Andersonbook,Olver:1993} for
reviews). We will use the odd operator
\begin{equation}
  \label{eq:3}
  \ddl{}{{}^*e^{a}},
\end{equation}
satisfying
\begin{equation}
  \label{eq:4}
  [\ddl{}{{}^*e^{a}},\ddl{}{{}^*e^{a}}]=0=[{}^*e^{a},{}^*e^{b}],\quad [\ddl{}{{}^*e^{a}},{}^*e^{b}]=\delta^b_a,
\end{equation}
where $[\cdot,\cdot]$ denotes the graded commutator, and thus for the
odd variables above the anti-commutator. In these terms, the
differential $d$ acts from the left as
\begin{equation}
  d={}^*e^{a}\partial_a-\half {D^c}_{ab}{}^*e^{a}{}^*e^{b
  }\ddl{}{{}^*e^{c}} \label{eq:17},
\end{equation}
or
\begin{equation}
  \label{eq:5}
   d: \omega_{a_1\dots a_k}\mapsto (k+1) \d_{[a_0}\omega_{a_1\dots
    a_k]}- \frac{k(k+1)}{2}{D^c}_{[a_0 a_1}\omega_{|c|a_2\dots a_k]}.
\end{equation}

\subsection{Hodge dual and co-differential}
\label{sec:hodge-dual-co}

We also assume that there is a pseudo-Riemannian metric,
\begin{equation}
g_{\mu\nu}={e^a}_\mu \eta_{ab} {e^b}_\nu\label{eq:11},
\end{equation}
where $\eta_{ab}$ is constant.
As usual, tangent space indices $a,b,\dots$ and world indices
$\mu,\nu,\dots$ are lowered and raised with $\eta_{ab}$, $g_{\mu\nu}$,
and their inverses $\eta^{ab},g^{\mu\nu}$.

We take $\epsilon_{a_1\dots a_n}=\epsilon_{[a_1\dots a_n]}$ completely
antisymmetric with $\epsilon_{1\dots n}=1$. The Hodge dual can then be
defined as the operator acting from the right,
\begin{equation}
  \star =\sum_{k=0}^n\frac{1}{k!(n-k)!} \ddl{{}^R}{{}^*e^{a_k}}\dots \ddl{{}^R}{{}^*e^{a_1}}
  {\epsilon^{a_1\dots a_k}}_{b_{k+1}\dots b_n}{}^*e^{b_{k+1}}\dots
{}^*e^{b_n},\label{eq:32}
\end{equation}
where $\ddl{{}^R}{{}^*e^{a}}$ is a derivative from the right.
In components, or in an abstract index notation, this gives
\begin{equation}
  \label{eq:22}
  \star: \omega_{a_1\dots a_k}\mapsto  \frac{1}{k!}\omega^{b_1\dots
    b_k}\epsilon_{b_1\dots b_ka_{k+1}\dots a_n}.
\end{equation}
It follows that
\begin{equation}
\star (\star \omega^k)=(-)^{t+k(n-k)}\omega^k,\quad
\ddl{}{{}^*e^{a}}(\star\omega)=\star (\omega {}^*e^{}_a), \label{eq:62}
\end{equation}
where
$(-)^t$ is the sign of ${\rm det}\,\eta_{ab}$, and, for a
variation,
\begin{equation}
  \label{eq:124}
\delta^V\star\omega=\star(\delta^V\omega)+(\delta^V{}^*e^{a})\star(\omega {}^*e_a).
\end{equation}

The operator acting from the
right
\begin{equation}
  \label{eq:36}
  \delta^R=\ddl{{}^R}{{}^*e^{a}} \d^a-\half  {}^*e^{a} {D_a}^{bc}\ddl{{}^R}{{}^*e^{c}}\ddl{{}^R}{{}^*e^{b}},
\end{equation}
or
\begin{equation}
  \label{eq:38}
  \delta^R: \omega_{a_1\dots a_k}\mapsto \d^{a_k}\omega_{a_1\dots a_{k-1}
    a_k}-\omega_{a_1\dots a_{k-1} a_k}{D_b}^{a_kb}
  -\frac{k-1}{2}\omega_{[a_1a_2\dots a_{k-2}|bc|} {D_{a_{k-1}]}}^{bc},
\end{equation}
satisfies
\begin{equation}
  \label{eq:34}
  d(\star\omega)=\star(\delta^R\omega).
\end{equation}
It is related to the standard co-differential $\delta^L$ acting from the
left through $\delta^R\omega^k=(-)^k\delta^L\omega^k$, with
\begin{multline}
\delta^L=-[\ddl{}{{}^*e^{a}}
\d^a-\half {D_a}^{bc}\ddl{}{{}^*e^{b}}\ddl{}{{}^*e^{c}} {}^*e^{a}]\\=-[\ddl{}{{}^*e^{a}}
\d^a-\half {D_a}^{bc} {}^*e^{a}\ddl{}{{}^*e^{b}}\ddl{}{{}^*e^{c}}-{D_c}^{bc}\ddl{}{{}^*e^{b}}].\label{eq:41}
\end{multline}

\subsection{Covariant expressions for (co-)differential}
\label{sec:covar-expr-co}

When there exists an affine Lorentz connection, with curvature and torsion
defined as in section \ref{sec:cartannonholo}, one may write
\begin{equation}
  d={}^*e^{a}D_a+\half {T^c}_{ab}{}^*e^{a}{}^*e^{b
  }\ddl{}{{}^*e^{c}} \label{eq:17a},
\end{equation}
or
\begin{equation}
  \label{eq:5a}
   d: \omega_{a_1\dots a_k}\mapsto (k+1) D_{[a_0}\omega_{a_1\dots
    a_k]}+ \frac{k(k+1)}{2}{T^c}_{[a_0 a_1}\omega_{|c|a_2\dots a_k]},
\end{equation}
and also
\begin{equation}
  \label{eq:36a}
  \delta^R=\ddl{{}^R}{{}^*e^{a}} D^a+\half  {}^*e^{a} {T_a}^{bc}\ddl{{}^R}{{}^*e^{c}}\ddl{{}^R}{{}^*e^{b}},
\end{equation}
or
\begin{equation}
  \label{eq:38a}
  \delta^R: \omega_{a_1\dots a_k}\mapsto D^{a_k}\omega_{a_1\dots a_{k-1}
    a_k}+\omega_{a_1\dots a_{k-1} a_k}{T_b}^{a_kb}
  +\frac{k-1}{2}\omega_{[a_1a_2\dots a_{k-2}|bc|} {T_{a_{k-1}]}}^{bc}.
\end{equation}
Finally,
\begin{equation}
  \delta^L=-[\ddl{}{{}^*e^{a}}
D^a+{T_c}^{bc}\ddl{}{{}^*e^{b}}+\half {T_a}^{bc}{}^*e^{a}\ddl{}{{}^*e^{b}}\ddl{}{{}^*e^{c}} ].\label{eq:41a}
\end{equation}
In components, $\delta$ is given by \eqref{eq:38a}, with an additional
overall sign of $(-)^k$.

In particular, for our purpose, it is convenient to write $n$, $n-1$
and $n-2$-forms in terms of duals of $0$, $1$ and $2$-forms,
$\omega^n=\star f=\mathbf e f dx^0\dots dx^{n-1}$, with $\mathbf
e={\rm det}\ {e^a}_\mu$,
\begin{equation}
  \omega^{n-1}=\star (j_a {}^*e^{a})\Longrightarrow d\omega^{n-1}=\star
    (D_a j^a+{T^b}_{ab} j^a), \label{eq:40}
\end{equation}
\begin{equation}
  \omega^{n-2} =\star (\half k_{ab}
{}^*e^{a}{}^*e^{b})\Longrightarrow d\omega^{n-2}=\star \big[(D_b {k_a}^b +{k_a}^b
{T^c}_{bc}+\half k_{bc}{T_{a}}^{bc}){}^*e^{a}\big],\label{eq:39}
\end{equation}
and to use covariant ``integration by parts'' inside $n$-forms,
\begin{multline}
   \label{eq:46} \star (v^{ab_1\dots b_m} D_a w_{b_1\dots b_m})=d[
\star(v^{ab_1\dots b_m}w_{b_1\dots b_m}{}^*e_a)] \\-\star[
(D_a+{T^c}_{ac})v^{ab_1\dots b_m} w_{b_1\dots b_m}].
\end{multline}

\section{Homotopy operators for the Euler-Lagrange complex}
\label{sec:homot-oper-euler}

On account of the (global) exactness of the horizontal part of the
variational bicomplex in vertical degree $1$, the variation of any
local form can be decomposed in terms of local forms as
\begin{equation}
  \label{eq:48}
  \begin{split}
    \delta^V(\star\omega^0) & =\varphi^i\vddl{[\delta^V
      (\star \omega^0)]}{\varphi^i}+d (\cI^{n}_\varphi[\delta^V(\star\omega^0)]),\\
  \delta^V(\star \omega^k)& =d (\cI^{n-k}_{\varphi} [\delta^V(\star \omega^k)])+
  \cI^{n-k+1}_{\varphi}(d [\delta^V(\star \omega^k)]), \quad {\rm for}\quad  k>0,
\end{split}
\end{equation}
for suitably defined ``homotopy'' operators of the variational
bi-complex,
\begin{equation}
  \label{eq:54}
  \cI^{n-k}_\varphi[\delta^V(\star \omega^k)]=\sum_{l=0}\frac{l+1}{k+l+1}
  \d_{\lambda_1\dots\lambda_l}
  (\varphi^i\vddl{}{(\d_{\lambda_1}\dots\d_{\lambda_l}\d_\rho\varphi^i)}
  {e^a}_\rho\ddl{[\delta^V(\star\omega^k)]}{{}^*e^{a}}),
\end{equation}
where $\vddl{}{(\d_{\lambda_1}\dots\d_{\lambda_l}\d_\rho\varphi^i)}$
are higher order Euler-Lagrange derivatives, see
e.g.~\cite{Andersonbook,Anderson1991,Olver:1993} for more details.

In order to simplify computations, note that
\begin{equation}
  \label{eq:56}
  \vddl{[\delta^V(\star
    \omega^0)]}{\varphi^i}=\vddl{(\star\omega^0)}{\phi^i},\quad
  \cI^{n-k}_\varphi[\delta^V(\star \omega^k)]=I^{n-k}_\varphi(\star \omega^k),
\end{equation}
with
\begin{equation}
  \label{eq:64}
  I^{n-k}_\varphi(\star \omega^k) = \sum_{l=0}\frac{l+1}{k+l+1}
  \d_{\lambda_1\dots\lambda_l}
  (\varphi^i\vddl{}{(\d_{\lambda_1}\dots\d_{\lambda_l}\d_\rho\phi^i)}{e^a}_\rho\ddl{[(\star\omega^k)]}{{}^*e^{a}}).
\end{equation}
Note also that, if $\omega^k_1$ is of first order in derivatives, this
simplifies to
\begin{equation}
  \label{eq:63}
  I^{n-k}_\varphi(\star
  \omega^k_1)=\star[\frac{1}{k+1}\varphi^i\ddl{\omega^k_1}{\d_a\phi^i}{}^*e_a].
\end{equation}

In order to prove equation \eqref{eq:74}, note that
$\cI^{n-1}_\varphi(\star S_{\delta^Vf})$ produces on-shell vanishing
terms for the full theory,
\begin{equation}
\cI^{n-1}_\varphi(\star S_{\delta^Vf})\approx 0\label{eq:67}.
\end{equation}
It follows that
\begin{equation}
  \label{eq:68}
  \delta^V(\star S_f)-\star S_{\delta^V f}=d (\star
  k_f)+\cI^{n}_\varphi(d[\delta^V (\star S_f)-\star S_{\delta^V f}]),
\end{equation}
with
\begin{equation}
  \label{eq:69}
  \star k_f=\cI^{n-1}_\varphi[\delta^V(\star S_f)-\star S_{\delta^V
    f}]=I^{n-1}_\varphi[\star S_{f(x)}]|_{f(x)=f}.
\end{equation}

Finally, for first order equations of motion, the breaking defined in
\eqref{eq:65} reduces to
\begin{equation}
  \label{eq:75}
  b_f[\varphi,\phi]=-\delta_f \phi^i \varphi^j \frac{\p}{\p\p_a\phi^j}\frac{\delta
    L}{\delta\phi^i}{}^*e_a.
\end{equation}

\section{Newman-Unti solution space}
\label{NP solution}

When conditions \eqref{gauge conditions} supplemented by the fall-off
conditions \eqref{fall-off conditions} are imposed, the asymptotic
expansion of on-shell spin coefficients, tetrads and the associated
components of the Weyl tensor can be determined. All the coefficients
in the expansions are functions of the three coordinates
$u,\zeta, \bar{\zeta}$. In this approach to the characteristic initial
value problem, freely specifiable initial data at fixed $u_0$ is given
by $\Psi_0 (u_0, r, \zeta, \bar{\zeta})$ in the bulk with the
fall-offs given below and by
$(\Psi^0_2 + \bar{\Psi}^0_2)(u_0, \zeta, \bar{\zeta})$,
$\Psi^0_1 (u_0, \zeta, \bar{\zeta})$ at $\mathscr{I}^+$. The
asymptotic shear $\sigma^0(u, \zeta, \bar{\zeta})$ and the conformal
factor $P(u, \zeta, \bar{\zeta})$ are free data at $\mathscr{I}^+$ for
all $u$.

Explicitly,
\begin{equation*}
\begin{split}
&\Psi_0=\frac{\Psi_0^0}{r^5} + \frac{\Psi_0^1}{r^6} + \frac{\Psi_0^2}{r^7} + \cO(r^{-8}),\\
&\Psi_1=\frac{\Psi_1^0}{r^4}-\frac{\xbar \eth
  \Psi_0^0}{r^5}+\frac{2\sigma^0\xbar\sigma^0\Psi_1^0 + \frac52 \eth\xbar\sigma^0 \Psi_0^0 + \frac12 \xbar\sigma^0 \eth \Psi_0^0 - \frac12 \xbar\eth \Psi_0^1}{r^6} + \cO(r^{-7}),\\
&\Psi_2=\frac{\Psi_2^0}{r^3}-\frac{\xbar \eth \Psi_1^0}{r^4} + \frac{2\eth\xbar\sigma^0 + \frac12 \lambda^0\Psi_0^0 + \frac32 \sigma^0\xbar\sigma^0\Psi_2^0 + \frac12 \xbar\sigma^0\eth\Psi_1^9 + \frac12 \xbar\eth^2\Psi_0^0}{r^5}+ \cO(r^{-6}),\\
&\Psi_3=\frac{\Psi_3^0}{r^2}-\frac{\xbar
  \eth\Psi_2^0}{r^3}+\cO(r^{-4}),\quad
\Psi_4=\frac{\Psi_4^0}{r}-\frac{\xbar \eth \Psi_3^0}{r^2}+\cO(r^{-3}),\\
\end{split}
\end{equation*}
\begin{equation*}
\begin{split}
&\rho=-\frac{1}{r}-\frac{\sigma^0\xbar\sigma^0}{r^3}+\cO(r^{-5}),\
\sigma=\frac{\sigma^0}{r^2}+ \frac{\bar{\sigma}^0 \sigma^0 \sigma^0 - \frac{1}{2} \Psi^0_0}{r^4}+\cO(r^{-5}),\\
&\tau=-\frac{\Psi^0_1}{2r^3}+ \frac{\frac{1}{2} \sigma^0
  \bar{\Psi}^0_1 + \bar{\eth} \Psi^0_0}{3r^4}+ \cO(r^{-5}),\
\alpha=\frac{\alpha^0}{r} +\frac{\xbar\sigma^0\xbar\alpha^0}{r^2}
+\frac{\sigma^0\xbar\sigma^0\alpha^0}{r^3}+\cO(r^{-4}),
\\
&\beta=-\frac{\xbar\alpha^0}{r}-\frac{\sigma^0\alpha^0}{r^2}
-\frac{\sigma^0\xbar\sigma^0\xbar\alpha^0+\half
  \Psi^0_1}{r^3}+\cO(r^{-4}),\
\gamma=\gamma^0-\frac{\Psi^0_2}{2r^2}+\frac{2 \bar{\eth} \Psi^0_1 +
  \alpha^0 \Psi^0_1 - \bar{\alpha}^0
  \bar{\Psi}^0_1}{6r^3}+\cO(r^{-4}), \\
&
 \mu=\frac{\mu^0}{r} -
 \frac{\sigma^0\lambda^0+\Psi^0_2}{r^2}+\frac{\sigma^0 \bar{\sigma}^0
   \mu^0 + \frac{1}{2}\bar{\eth} \Psi^0_1}{r^3}+\cO(r^{-4}),\
 \nu=\nu^0-\frac{\Psi^0_3}{r}+\frac{\xbar \eth
  \Psi^0_2}{2r^2}+\cO(r^{-3}),\\
&
\lambda=\frac{\lambda^0}{r}-\frac{\xbar\sigma^0
  \mu^0}{r^2}+\frac{\sigma^0 \bar{\sigma}^0\lambda^0 + \frac{1}{2}
  \bar{\sigma}^0 \Psi^0_2}{r^3}+\cO(r^{-4}),\\
\end{split}
\end{equation*}
\begin{equation*}
\begin{split}
&X^\zeta = \overline{X^{\bar{\zeta}}}= \frac{\bar{P}\Psi^0_1}{6 r^3}
+\cO(r^{-4}), \quad
\omega=\frac{\xbar \eth \sigma^0}{r} -\frac{\sigma^0\eth \xbar\sigma^0
  +\half \Psi^0_1}{r^2}+\cO(r^{-3}),\\
&U=-r(\gamma^0+\xbar\gamma^0) + \mu^0-\frac{\Psi^0_2 + \xbar
  \Psi^0_2}{2r}+ \frac{\bar{\eth} \Psi^0_1 + \eth
  \bar{\Psi}^0_1}{6r^2}+ \cO(r^{-3}),\\
&L^\zeta=\overline{{\bar{L}}^{\bar{\zeta}}}=-\frac{\sigma^0
  \bP}{r^2}+\cO(r^{-4}), \quad
L^{\bar{\zeta}}=\overline{{\bar{L}}^\zeta}=\frac{P}{r}+\frac{\sigma^0
  \xbar\sigma^0 P}{r^3}+\cO(r^{-4}),\\
\end{split}
\end{equation*} where
\begin{equation*}
\begin{split}
&\alpha^0=\half \bP \p \ln P,\quad
\gamma^0=-\half \p_u \ln \bP,\quad \nu^0=\xbar \eth (\gamma^0+\xbar\gamma^0),\\
&\mu^0=-\half P \bP \p\xbar\p \ln P\bP = -\half \bar{\eth}\eth \ln
P\bP= -\frac{R}{4} ,\quad
\lambda^0= \dot{\xbar\sigma^0} + \xbar \sigma^0 (3\gamma^0 - \xbar \gamma^0),\\
&\Psi^0_2 - \bar{\Psi}^0_2 = \bar{\eth}^2 \sigma^0 - \eth^2
\bar{\sigma}^0 + \bar{\sigma}^0 \bar{\lambda}^0 - \sigma^0 \lambda^0
\\
&\Psi^0_3 = - \eth \lambda^0 + \bar{\eth} \mu^0 , \\
&\Psi^0_4 = \bar{\eth} \nu^0 - (\partial_u + 4 \gamma^0) \lambda^0,
\end{split}
\end{equation*}
and
\begin{align*}
&\p_u\Psi^0_0 + (\gamma^0 + 5 \xbar \gamma^0)\Psi^0_0=\eth\Psi^0_1+3\sigma^0\Psi^0_2,\\
&\p_u\Psi^0_1 + 2 (\gamma^0 + 2 \xbar \gamma^0)\Psi^0_1=\eth\Psi^0_2+2\sigma^0\Psi^0_3,\\
&\p_u\Psi^0_2 + 3 (\gamma^0 + \xbar \gamma^0)\Psi^0_2=\eth\Psi^0_3 + \sigma^0\Psi^0_4,\\
&\p_u\Psi^0_3 + 2 (2 \gamma^0 + \xbar \gamma^0)\Psi^0_3=\eth\Psi^0_4,\\
&\p_u
 \mu^0=-2(\gamma^0+\xbar\gamma^0)\mu^0+\xbar\eth\eth(\gamma^0+\xbar\gamma^0),\\
&\p_u\alpha^0=-2\gamma^0\alpha^0-\xbar\eth\xbar\gamma^0,
\end{align*}
\begin{align*}
\p_u \Psi_0^1 + (2\gamma^0 + 6\xbar\gamma^0) \Psi_0^1 = - \xbar\eth (\eth\Psi_0^0 + 4 \sigma^0 \Psi_1^0),
\end{align*}
\begin{multline*}
\p_u \Psi_0^2 + (3\gamma^0 + 7\xbar\gamma^0) \Psi_0^2 = -
\frac12\xbar\eth \eth\Psi_0^1 + 3 \mu^0 \Psi_0^1 + 5 (\Psi_1^0
\Psi_1^0 - \Psi_0^0 \Psi_2^0 - \frac12 \Psi_0^0 \xbar\Psi_2^0)\\
+ 5 \xbar\eth \sigma^0 \xbar\eth\Psi_0^0 + 3
\eth\xbar\sigma^0\eth\Psi_0^0 + \frac52 \sigma^0\xbar\eth^2\Psi_0^0 +
\frac52 \eth^2\xbar\sigma^0 \Psi_0^0 + \frac12 \xbar\sigma^0
\eth^2\Psi_0^0 + \frac92 \sigma^0\xbar\sigma^0 \eth\Psi_1^0\\
+12\sigma^0\eth\xbar\sigma^0 \Psi_1^0 +
2\xbar\sigma^0\eth\sigma^0\Psi_1^0 +
\frac{15}{2}\xbar\sigma^0(\sigma^0)^2\Psi_2^0 +
\frac52\sigma^0\lambda^0\Psi_0^0.
\end{multline*}

\section{Parameters of residual gauge transformations}\label{ASG}

For computational purposes, it turns out to be more convenient to
determine the parameters of residual gauge transformations by using
the generating set given in \eqref{eq:15} rather than the one in
\eqref{eq:23}.

Asking that conditions \eqref{gauge conditions} be preserved on-shell yields
\begin{itemize}
\item
  $0=\delta_{\xi',\omega'}\; e_1^u=-\p_r \xi'^u \Longrightarrow
  \xi'^u=f(u,\zeta,\bar \zeta)$.
\item
  $0=\delta_{\xi',\omega'}\; e_2^u=-e_2^\alpha \p_\alpha f + \omega'^{12}
  \Longrightarrow \omega'^{12}=\p_u f + X^A \p_A f$.
\item
  $0=\delta_{\xi',\omega'}\; e_3^u=-e_3^\alpha \p_\alpha f + \omega'^{42}
  \Longrightarrow \omega'^{24}= L^A \p_A f$.
\item
  $0=\delta_{\xi',\omega'}\; e_4^u=-e_4^\alpha \p_\alpha f + \omega'^{32}
  \Longrightarrow \omega'^{23}= \bL^A \p_A f$.
\item
  $0=\delta_{\xi',\omega'}\; e_1^r=-e_1^\alpha \p_\alpha \xi'^r +
  \omega'^{2a}e_a^r \Longrightarrow \xi'^r=-\p_u f r + Z(u,\zeta,\bar \zeta) - \p_A
  f \int^{+\infty}_r dr[\omega \bL^A + \bomega L^A + X^A]$.
\item
  $0=\delta_{\xi',\omega'}\; e_1^A=-e_1^\alpha \p_\alpha \xi'^A +
  \omega'^{2a}e_a^A \Longrightarrow \xi'^A=Y^A(u,\zeta,\bar \zeta) - \p_B f
  \int^{+\infty}_r dr[L^A \bL^B + \bL^AL^B ]$.
\item
  $\delta_{\xi',\omega'}\;\bar\pi=0\iff 0=\delta_{\xi',\omega'}\;
  \Gamma_{321}=l^\mu \p_\mu \omega'^{41} + \Gamma_{32a} \omega'^{2a}
  \Longrightarrow \omega'^{14}=\omega'^{14}_0(u,\zeta,\bar \zeta) + \p_A f
  \int^{+\infty}_r dr[\bar{\lambda} \bL^A + \bar{\mu} L^A]$.
\item $\delta_{\xi',\omega'}\;\pi=0\iff 0=\delta_{\xi',\omega'}\;
  \Gamma_{421}=l^\mu \p_\mu \omega'^{31} + \Gamma_{42a} \omega'^{2a}
  \Longrightarrow
\omega'^{13}=\omega'^{13}_0(u,\zeta,\bar \zeta) + \p_A f \int^{+\infty}_r
dr[\lambda L^A + \mu \bL^A]$.
\item
  $\delta_{\xi',\omega'}\;(\epsilon-\bar\epsilon)=0\iff
  0=\delta_{\xi',\omega'}\; \Gamma_{431}=l^\mu \p_\mu \omega'^{43} +
  \Gamma_{43a} \omega'^{2a} \Longrightarrow
  \omega'^{34}=\omega'^{34}_0(u,\zeta,\bar \zeta) - \p_A f \int^{+\infty}_r
  dr[(\bar{\alpha}-\beta) \bL^A + (\bar{\beta}-\alpha) L^A]$.

\item $\epsilon+\bar\epsilon=0=\kappa=\bar\kappa$ is equivalent to
  $\Gamma_{211}= \Gamma_{311}=\Gamma_{411}=0$, $\rho-\bar\rho=0$ is
  equivalent to $\Gamma_{314}-\Gamma_{413}=0$ while
  $\tau-\bar\alpha-\beta=0$ is equivalent to
  $\Gamma_{213}-\Gamma_{312}=0$. On-shell, i.e., in the absence of
  torsion, these conditions on spin coefficients hold as a consequence
  of the tetrad conditions imposed in \eqref{gauge conditions}. It
  follows that requiring these conditions to be preserved on-shell by
  gauge transformations does not give rise to new conditions on the
  parameters. This can also be checked by direct computation.
\end{itemize}
Asking that the fall-off conditions \eqref{fall-off conditions} be
preserved on-shell yields
\begin{itemize}
\item $\delta_{\xi',\omega'}\; e_2^A=\cO(r^{-1}) \Longrightarrow \p_u Y^A=0$.
\item $\delta_{\xi',\omega'}\; g_{\zeta\zeta}=\cO(r^{-1})
  \Longrightarrow \bar{\partial} Y^{\zeta}=0\iff Y^\zeta = Y(\zeta)$.
\item $\delta_{\xi',\omega'}\; g_{\bar{\zeta}\bar{\zeta}}=\cO(r^{-1})
  \Longrightarrow \partial Y^{\bar{\zeta}}=0\iff Y^{\bar{\zeta}} = \bar{Y} (\bar{\zeta})$.
\item
  $\delta_{\xi',\omega'} \; \Gamma_{314} = \mathcal{O}(r^{-3})
  \Longrightarrow Z =\half \bDelta f$.
\item
  $\delta_{\xi',\omega'} \; \Gamma_{312} =\mathcal{O}(r^{-2})
  \Longrightarrow \omega'^{14}_0 = (\gamma^0 + \bar{\gamma}^0)P
  \bar{\partial} f - P \partial_u \bar{\partial}f$.
\item
  $\delta_{\xi',\omega'} \; \Gamma_{412} =\mathcal{O}(r^{-2})
  \Longrightarrow \omega'^{13}_0 = (\gamma^0 + \bar{\gamma}^0)\bar{P}
  {\partial} f - \bar{P} \partial_u {\partial}f$.
\item $\delta_{\xi',\omega'}\; \Psi_0 = \mathcal{O}(r^{-5})$ does not
  impose further constraints.
\end{itemize}

\section{Action on solution space: original parametrization}
\label{sec:oracti}

Besides \eqref{conformal factors}, if $s_o=(Y,\bar Y,f,\omega'_0)$, one finds
\begin{equation}
\begin{split}
&-\delta_{s_o} \sigma^0=[Y \p +\bY \bp + f \p_u + \p_u f +
2\omega'^{34}_0]\sigma^0 - \eth^2 f,\\
&-\delta_{s_o} \Psi^0_0=[Y \p +\bY \bp + f \p_u + 3\p_u f +
2\omega'^{34}_0]\Psi^0_0 + 4 \Psi^0_1 \eth f,\\
&-\delta_{s_o} \Psi^0_1=[Y \p +\bY \bp + f \p_u + 3\p_u f +
\omega'^{34}_0]\Psi^0_1 + 3 \Psi^0_2 \eth f,\\
&\hspace{-1cm} -\delta_{s_o} \left(\frac{\Psi^0_2 +
    \bar{\Psi}^0_2}{2}\right)=[Y \p +\bY \bp + f \p_u + 3\p_u f
]\left(\frac{\Psi^0_2 + \bar{\Psi}^0_2}{2}\right)  \\ &\hspace{9cm}
+  \Psi^0_3 \eth f +\bar{\Psi}^0_3 \overline{\eth} f.
\end{split}
\label{transfo solution space 1}
\end{equation}
When $\Psi_0$ can be expanded in powers of $1/r$, $
\Psi_0 = \sum_{n=0}^\infty \frac{\Psi_0^n}{r^{n+5}}$,
one also has
\begin{multline}\label{transfo solution space 2}
-\delta_{s_o} \Psi^1_0 = [Y \p +\bY \bp + f \p_u + 4\p_u f + 2\omega'^{34}_0] \Psi^1_0
\\ + [- \frac{5}{2} \bDelta f - 5 \eth f \overline{\eth} - \overline{\eth}f \eth] \Psi_0^0
- 4 \sigma^0 \overline{\eth} f \Psi^0_1,
\end{multline}
\begin{multline}\label{transfo solution space 3}
-\delta_{s_o} \Psi^2_0 = [Y \p +\bY \bp + f \p_u + 5\p_u f
+ 2\omega'^{34}_0] \Psi^2_0
+ [- 3 \bDelta f -  3\eth f \overline{\eth} - \overline{\eth}f \eth] \Psi_0^1
\\ + [5 \overline{\eth} \sigma^0 \overline{\eth}f + 15 \eth
\overline{\sigma}^0 \eth f+ 5 \sigma^0 \overline{\eth}f
\overline{\eth} + 3 \bar{\sigma}^0 \eth f \eth] \Psi^0_0
+ 12 \sigma^0 {\bar{\sigma}}^0 \eth f \Psi^0_1.
\end{multline}
By induction, we deduce
\begin{multline}\label{transfo solution space 4}
  -\delta_{s_o} \Psi^n_0 = [Y \p +\bY \bp + f \p_u + (n+3)\p_u f + 2\omega'^{34}_0] \Psi^n_0
\\+ (\text{inhomogeneous terms}).
\end{multline}
For later purposes, we also give the variations of composite
quantities in terms of free data,
\begin{equation}
\begin{split}
&-\delta_{s_o} \lambda^0=[Y \p +\bY \bp + f \p_u +  2\p_u f
- 2\omega'^{34}_0]\lambda^0 - \p_u\xbar\eth^2 f + (\xbar \gamma^0 - 3
\gamma^0) \xbar\eth^2 f,\\
&-\delta_{s_o} \Psi^0_2=[Y \p +\bY \bp + f \p_u + 3\p_u f
]\Psi^0_2 + 2 \Psi^0_3 \eth f,\\
&-\delta_{s_o} \Psi^0_3=[Y \p +\bY \bp + f \p_u + 3\p_u f -
\omega'^{34}_0]\Psi^0_3 + \Psi^0_4 \eth f,\\
&-\delta_{s_o} \Psi^0_4=[Y \p +\bY \bp + f \p_u + 3\p_u f -
2\omega'^{34}_0]\Psi^0_4.
\end{split}
\label{transfo solution space}
\end{equation}

\section{Useful relations}
\label{Useful relations}

Some useful relations for the computation of the current algebra are
summarized here.
\begin{align*}
&\p_u f=\frac12 (\eth\cY+\xbar\eth\xbar\cY) + f(\gamma^0+\xbar\gamma^0),\\
&\hat f =\frac12 f_1 (\eth\cY_2 + \xbar\eth\xbar\cY_2) + \cY_1 \eth
  f_2 + \xbar\cY_1 \xbar\eth f_2 - (1\leftrightarrow2),\\
&\hat \cY = \cY_1 \eth^2 \cY_2 - \cY_2 \eth^2 \cY_1,\quad \hat
  {\xbar\cY} = \xbar\cY_1 \xbar\eth^2 \xbar\cY_2 - \xbar\cY_2
    \xbar\eth^2 \xbar\cY_1,\\
  &\eth^2\hat\cY=\eth\cY_1 \eth^2\cY_2 + \cY_1 \eth^3\cY_2
    - (1\leftrightarrow2),\quad \eth\xbar\eth\hat{\xbar\cY}
    =\xbar\cY_1\eth\xbar\eth^2\xbar\cY_2 - (1\leftrightarrow2),\\
  &\eth^3\hat\cY=2\eth\cY_1 \eth^3\cY_2 + \cY_1 \eth^4\cY_2
    - (1\leftrightarrow2),\quad \eth^2\xbar\eth\hat{\xbar\cY}
    =\xbar\cY_1\eth^2\xbar\eth^2\xbar\cY_2 - (1\leftrightarrow2),\\
  &\xbar\eth\eth^3\cY=2\cY\eth^2\mu^0
    + 4\eth\mu^0\eth\cY,\quad\xbar\eth^2\eth^2\cY=2\xbar\eth\eth\mu^0
    \cY
    + 2\xbar\eth\mu^0\eth\cY + 4(\mu^0)^2\cY, \\
&\eth\hat f=\frac12 f_1 \eth(\eth\cY_2+\xbar\eth\xbar\cY_2) + \cY_1
   \eth^2 f_2 + \xbar\cY_1\eth\xbar\eth f_2 + \frac12(\eth\cY_1
 - \xbar\eth\xbar\cY_1)\eth f_2 - (1\leftrightarrow2),\\
&\eth\xbar\eth \xbar\cY=2\mu^0\xbar\cY,\quad
  \xbar\eth\eth\cY=2\mu^0\cY,
  \quad \p_u\eth\cY=2\xbar\nu^0\cY,\\
  &\p_u \eth f=\frac12\eth(\eth\cY+\xbar\eth\xbar\cY)
    +\eth f(\gamma^0-\xbar\gamma^0)+f \xbar\nu^0,\\
&\p_u\eth^2\cY=2\eth\xbar\nu^0\cY + 2 \xbar\nu^0\eth\cY - 2\xbar\gamma^0\eth^2\cY,\\
  &\p_u\eth\xbar\eth\xbar\cY=2\eth\nu^0\xbar\cY
    - 2 \xbar\gamma^0\eth\xbar\eth\xbar\cY,\\
  &\p_u\eth^2 f=\frac12\eth^2(\eth\cY+\xbar\eth\xbar\cY)
    + \eth^2 f (\gamma^0-3\xbar\gamma^0) + f\eth\xbar\nu^0,\\
  &\p_u\eth\xbar\eth f=\frac12\eth\xbar\eth(\eth\cY+\xbar\eth\xbar\cY)
    - \eth\xbar\eth f (\gamma^0 + \xbar\gamma^0)
    + \xbar\eth f \xbar\nu^0   + \eth f \nu^0 + f\eth\nu^0,\\
  &\p_u\eth\xbar\sigma^0=\eth\lambda^0 + \xbar\nu^0\xbar\sigma^0
    - (\xbar\gamma^0+3\gamma^0)\eth\xbar\sigma^0,\\
  &\p_u \eth\mu^0=\xbar\eth\xbar\nu^0 - 2\mu^0\xbar\nu^0
    - 2(\gamma^0+2\xbar\gamma^0)\eth\mu^0,\\
&\xbar\eth\eth\xbar\nu^0=\eth^2\nu^0-2\mu^0\xbar\nu^0.
\end{align*}
In case one wants to compute the current algebra from the expressions
derived in the standard Cartan formalism \cite{Barnich:2016rwk},
one needs to transform the spin coefficients into a Lorentz
connection with a space-time index in NU gauge. Using
the notations of subsection
\ref{sec:relat-newm-penr}, together with the gauge choice for the
tetrads \eqref{gauge
  conditions} (and thus also
\eqref{cotetrad 4d case}), we have
\begin{align*}
  \Gamma_{12u} &= -(\gamma + \bar{\gamma}) -\tau  X^A \bar{L}_A  -
\bar{\tau} X^A L_A, &\Gamma_{12A} = \tau \bar{L}_A + \bar{\tau} L_A,\\
  \Gamma_{13u} &= - \tau - \sigma X^A \bar{L}_A - \rho X^A L_A,
&\Gamma_{13A} = \sigma \bar{L}_A + \rho L_A, \\
  \Gamma_{14u} &= -\bar{\tau} - \bar{\sigma} X^A L_A - \rho X^A
                 \bar{L}_A,
&\Gamma_{14A} = \rho \bar{L}_A + \bar{\sigma} L_A, \\
  \Gamma_{23u} &= \bar{\nu} + \bar{\lambda} X^A \bar{L}_A + \bar{\mu}
                 X^A L_A,
&\Gamma_{23A} = - \bar{\lambda} \bar{L}_A - \bar{\mu} L_A ,\\
  \Gamma_{24u} &= \nu + \mu X^A \bar{L}_A + \lambda X^A L_A , &\Gamma_{24A} = - \mu \bar{L}_A - \lambda L_A ,\\
  \Gamma_{34u} &= (\gamma - \bar{\gamma}) + (\beta - \bar{\alpha}) X^A
                 \bar{L}_A + ( \alpha - \bar{\beta} ) X^A L_A ,
&\Gamma_{34A} = (\bar{\alpha} - \beta ) \bar{L}_A + (\bar{\beta} -\alpha   ) L_A ,\\
  \Gamma_{abr} &= 0.
\end{align*}

\providecommand{\href}[2]{#2}\begingroup\raggedright\endgroup


\end{document}